%% file: ms.tex
\documentclass[12pt,preprint]{emulateapj}

\usepackage{natbib}

\shorttitle{Cannibalization and Rebirth in NGC5387}
\shortauthors{Beaton, Martinez-Delgado, et al. }

\begin{document}

\title{Cannibalization and Rebirth in the NGC5387 System. I. The Stellar Stream \& Star Forming Region\altaffilmark{*}}

\author{
Rachael L. Beaton\altaffilmark{1},
David Mart{\'{\i}}nez-Delgado\altaffilmark{2,3,4},
Elena D'Onghia\altaffilmark{5,6},
Stefano Zibetti\altaffilmark{7},
R. Jay Gabany\altaffilmark{8},
Kelsey E. Johnson\altaffilmark{1},
Steven R. Majewski\altaffilmark{1},
Michael Blanton\altaffilmark{9},
Anne Verbiscer\altaffilmark{1}
}

\altaffiltext{1}{Department of Astronomy, University of Virginia,
    Charlottesville, VA, USA: rbeaton@virginia.edu}
\altaffiltext{2}{Max Planck Institut fur Astronomie, Heidelberg, Germany}
\altaffiltext{3}{Astronomisches Rechen-Institut, Zentrum fur Astronomie de Universitat Heidelberg, Germany}
\altaffiltext{4}{Humbolt Fellow for Advanced Research}
\altaffiltext{5}{Department of Astronomy, University of Wisconsin---Madison, Madison, WI, USA}
\altaffiltext{6}{Alfred P. Sloan Fellow}
\altaffiltext{7}{INAF-Osservatorio Astrofisico di Arcetri, Firenze, Italy}
\altaffiltext{8}{Black Bird II Observatory, Alder Springs, CA, USA}
\altaffiltext{9}{Department of Physics, New York University, New York, NY, USA}
\altaffiltext{*}{ Based on observations with the VATT: the Alice P. Lennon Telescope and the Thomas J. Bannan Astrophysics Facility.}

\begin{abstract}
 We have identified a low surface brightness stellar stream 
  from visual inspection of SDSS imaging for the edge-on, spiral galaxy NGC\,5387. 
 A blue overdensity was also identified in SDSS coincident with the stream intersection with the NGC\,5387 disk.
 The overdensity was also detected in the GALEX Deep Imaging Survey
  and found to contribute 38\% of the total FUV integrated flux from NGC\,5387, 
  which suggests that the region is actively forming stars.
 Deeper imaging was acquired with the Vatican Advanced Technology Telescope (VATT) in the $B$, $V$, and $R$ filters
  that confirmed the presence of both the stellar stream and the blue overdensity.
 Analysis of the VATT photometry indicates the stellar stream is red in color, $B-V$ = 0.7,
  and has a stellar mass of 6$\times10^{8} M_{\odot}$, which implies a merger ratio of 1:50.
 Assessment of the stream morphology suggests that the merger event has a circular radius, R$_{circ}\sim$ 16 kpc, 
  the stream formed in $\sim$400 Myr, and the progenitor had a total mass of $\sim$2$\times$10$^{10}$ M$_{\odot}$. 
 Spectroscopy from LBT+MODS1 was used to determine that the blue overdensity is at the same redshift 
  as NGC\,5387, consists of young stellar populations ($\sim$10 Myr), is metal-poor (e.g.~$12 + log(O/H) = 8.03$), and 
  forming stars at an enhanced rate ($\sim$ 1-3 M$_{\odot}$ year$^{-1}$, depending on tracer) 
  given its total stellar mass (2 $\times$ 10$^{7}$ $M_{\odot}$).
 Several interpretations are posited to explain these observational data, 
  of which the most likely are (i) that the blue overdensity is a region of enhanced
 star formation in the outer disk of NGC\,5387 induced by the minor accretion event, and 
 (ii) that the blue overdensity is the progenitor of the stellar stream
  undergoing a period of enhanced star formation as a result of its interaction with NGC\,5387.
 Confirmation and theoretical exploration of these scenarios are presented in a companion paper.
\end{abstract}


\section{Introduction}\label{sec:intro}

In the hierarchical galaxy formation paradigm,
 the haloes of Milky Way (MW) sized galaxies are built through the accretion of less massive galaxies.
In Local Group, this process has been observed as streams of gaseous or stellar material
 in the Milky Way \citep[e.g.~ the Large Magellanic Cloud (LMC); ][]{nidever2010,nidever2013}, 
 Andromeda \citep[e.g.~M33; ][]{mcconnachie2013,wolfe2013},
 and in Milky-Way analogue galaxies across the Local Volume \citep[e.g.~][]{streamsurveypaper,putman2012}.

$N$-body modeling of minor accretion events within the $\Lambda$CDM context 
 indicate that streams can take on a variety of morphologies, which are determined by
 their orbital parameters \citep[See Figure 2]{kvj_2008}:
 ``great circles'' from circular orbits, ``umbrellas'' and ``shells'' from radial orbits,
 and ``mixed'' morphologies from old mergers (several Gyr) of various orbital types. 
Similar modeling that tracks the build-up of material in halos suggests
 that while all MW-type galaxies will show tidal debris in their haloes, 
 the detailed distribution of debris depends on the specific accretion history ---
 more specifically, the number, mass, and timing of individual accretions \citep{bj2005}.

In contrast to detailed predictions from simulations, 
 the observational portrait of minor accretion events is far from complete,
 owing primarily to the inherent difficulty of detecting low surface brightness tidal features. 
However, several observational analogues to the canonical \citet{kvj_2008} stream morphologies 
 have been discovered in the Local Volume \citep{streamsurveypaper}, 
 including the ``great circle'' morphology of NGC\,5907 
 \citep{shang1998,zheng1999,dmd2008} and Messier 63 \citep{chonis2011},
 the ``umbrella'' in NGC\,4651 \citep{streamsurveypaper}, and the
 mixed type structure in NGC\,1055 \citep{streamsurveypaper}. 
In addition to the three canonical structures in the \citet{kvj_2008} simulations,
 \citet{streamsurveypaper} identified additional stream morphologies, including ``spikes'' (NGC\,5866),
 partial disruptions (NGC\,4216), and ``giant plumes'' (NGC\,1084), that 
 are likely variations on the canonical forms due to the effect of viewing angle(s). 

Beyond these isolated discoveries, there exist few large scale searches for 
 tidal features that are sufficient to provide crucial feedback to simulations 
 on either the frequency of streams or the distribution of stream properties, including 
  stream morphology, debris mass, progenitor mass, or remnant mass.
Using morphological techniques to analyze a sample of 474 galaxies in the
 Sloan Digital Sky Survey (SDSS) DR7 \citep{sdss_dr7}, \citet{mbd2011} found that 6\% of galaxies exhibit
 clear tidal features and 19\% show faint features at a limiting surface brightness
 of 28 mag arcsec$^{-2}$.  
Along a similar vein, \citet{cfht_tidal} find 12\% of galaxies imaged in the wide-field 
 component of the Canada-France-Hawaii Telescope Legacy Survey exhibited
 tidal features, though this includes both major and minor merging events. 
These estimates of the frequency of tidal features are in stark contrast to those
 anticipated from the $\Lambda$-CDM paradigm, e.g., that {\it all} MW-sized 
 galaxies should have signatures of accretion events.
 The disparity is  largely due to the shallow surface brightness limit of imaging 
 surveys like SDSS --- most features in \citet{bj2005}
 have surface brightnesses exceeding 28 mag arcsec$^{-2}$.

Although large scale searches for tidal debris are limited, 
 the satellite galaxies of the Local Group (LG) --- those dwarf galaxies
 currently associated with the MW or with Andromeda --- do place some constraints
 on the process of satellite accretion. 
One significant constraint is the observed morphology-density relationship in the LG \citep{mateo1998}, 
 the simple observation that, with the exception of the Magellanic System 
 (the Large and Small Magellanic Cloud (LMC, SMC) and associated debris) and M\,33, 
 the early type dwarfs (predominantly old dwarf spheroidals, dSph) are predominantly discovered
 embedded within the halo of their host, whereas late type dwarfs 
 (star forming dwarf irregulars, dIrrs) are found in relative isolation
  or on their first passage around their host (e.g.~the LMC and SMC).
Beyond the LG, morphology-density relationships exist in other host-satellite systems \citep{t11,geha2012}.
\citet{gp2009} expanded the morphological dichotomy in the LG by noting that the majority of dwarf 
 galaxies within $\sim$270 kpc of their host are {\it undetected} in H\,I surveys, 
 meaning not only are satellites quenched for current star formation, 
 but they have no appreciable gas content to fuel future star formation.
Combined, the morphology-density relationship \citep{mateo1998} and the HI disparity \citep{gp2009} 
 imply that the minor merger process converts gas-rich, star-forming dIrrs into gas-devoid, quiescent dSphs,
 as suggested by, for example, \citet{grebel2003,ho2012}. 
Recent simulations also support this portrait \citep[e.g.~][]{elena2009,lokas2010}, though the competing
 roles of reionization, star formation feedback, and gas stripping remain difficult 
 to disentangle for individual dwarf galaxies in even the most advanced simulations.

Though the density-morphology relationship is compelling, several observational 
 studies suggest that the observed morphological change of satellite galaxies
 may not only be driven by interactions with a parent galaxy.
For example, studies of close ($<$ 50 kpc) satellite-host pairs in SDSS do not show strong morphological differences
 between a sample of ``isolated'' dwarfs and close-pair satellites \citep{phillips2013}.
In the LG, a number of dSph type galaxies have been discovered at large projected radii ($>$200 kpc) 
 or in a ``first in-fall'' scenario, 
 most notably the satellites, 
 AndXIV \citep{majewski2007}, AndXVIII, AndXXVIII \citep{slater2011} and AndXXIX \citep{bell2011}.
These ``isolated'' dSphs have had limited gravitational interaction with their parent system,
 suggesting that there must be another evolutionary path 
 capable of creating gas-devoid, dispersion supported galaxies.
Furthermore, the recent detection of a companion satellite to NGC\,4449 \citep{dmd2012}
 and implications of a companion satellite to IC\,10 \citep{nidever_ic10}, as well as more evidence of the
 complex structure of the dwarf-dwarf Magellanic system \citep{nidever2013}, 
 strongly suggest that dwarf galaxies, like their MW-sized hosts, may also have complex 
 minor-minor interaction histories that could strongly impact their overall evolution.

Despite challenges to the morphology-density relationship, the observational evidence on 
 satellite systems implies that the process of minor merging causes significant 
 morphological changes in the satellite. 
Since few ``in-tact'' star forming satellites are found embedded in their host ---
 and even so called ``transitional'' morphological types 
 \citep[e.g.~galaxies similar to Leo T][]{leotdisc,leotsfh,leotgas} are rare, 
 the gas removal and associated morphological changes proceed relatively quickly compared to the ``lifetime'' 
 of the average satellite in the host halo ($\sim$ several Gyrs for some dSphs).
Furthermore, the morphological change must occur by a means that neither completely disrupts the satellite
 given the large number of ``intact'' satellites in the LG, 
 nor often leaves clear signs of tidal stripping given that relatively few dSphs have significant tidal features.
Interestingly, \citet{kvj_2008} found that of 153 surviving satellites in their simulations, 
 only 21\% lost more than 1\% of their luminous matter but only 4\% of the streams 
 had still identifiable progenitors (e.g., bound halos). 
Thus, to understand fully the phase of minor merging during which gas is removed from
 satellites, it is necessary to identify mergers in the earliest stages of disruption.

In this paper, we present the case of the NGC\,5387 system --- 
 a spiral galaxy identified to be in the rare state of an ongoing satellite accretion event
 based on imaging from the Sloan Digital Sky Survey Data Release 9 \citep[SDSS-DR9 hereafter]{dr9_cat}. 
The event has two notable features, a stellar stream and blue overdensity.
In Section~\ref{sec:betterdata}, we present deep imaging of NGC\,5387 with the 
 Vatican Advanced Technology Telescope (VATT).
In Section~\ref{sec:spec}, we present spectroscopy of the blue overdensity to determine its detailed properties.
In Section~\ref{sec:interp}, several interpretations for the origin of the blue overdensity are presented and discussed. 
In a companion paper, numerical simulations are used to identify the most likely interpretation,
 and the implications of this system will be explored.

\begin{figure*}[tb]
\plotone{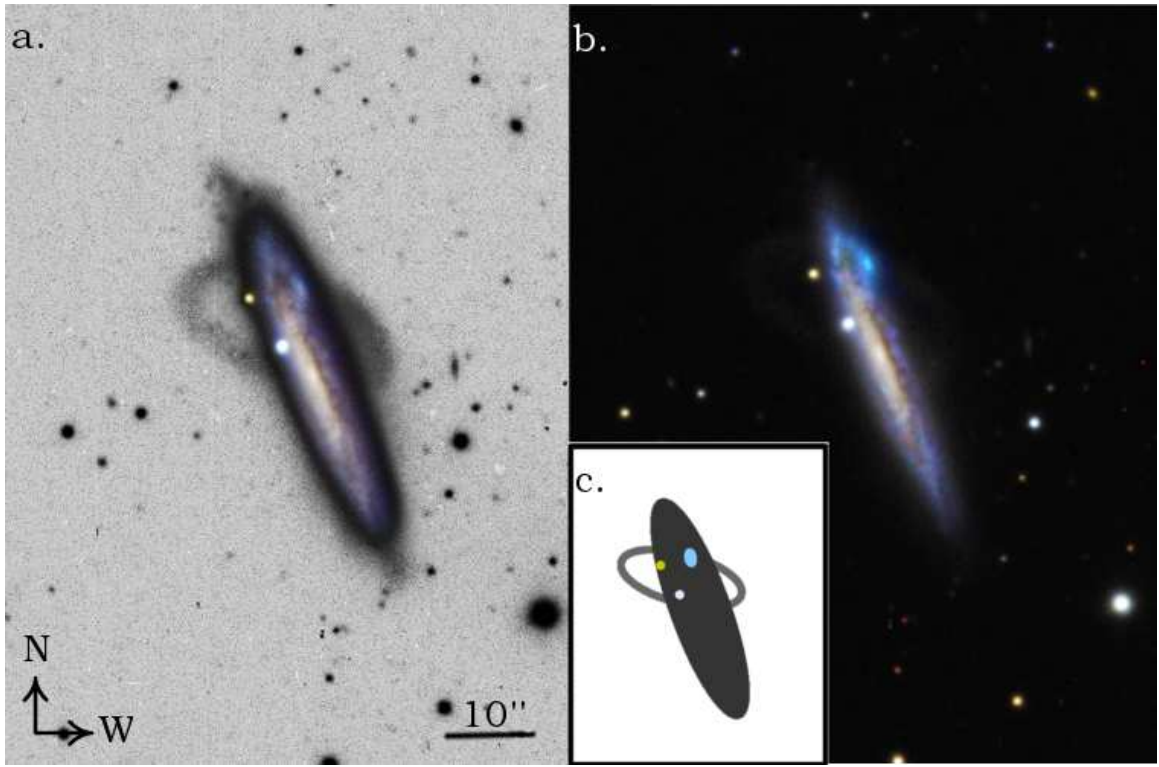}
\caption{\label{fig:prettyimage} SDSS color image of NGC\,5387; in panel a, 
 inset in the greyscale $R$ band image from VATT, 
 in panel b, merged with GALEX FUV and VATT imaging,
 and in panel c, a schematic view of the system.
 The color image reveals NGC\,5387 to have the features typical of a Milky Way type spiral galaxy,
  except for a small blue region on the north-western extent of the outer disk.
 The deep $R$ band image (panel a) highlights a narrow stellar stream extending from the north-eastern outer
  disk to the east, with a connection to another, broader feature on the western side of the disk.
 The bright circular regions on disk near the stream, a yellow-orange region and a blue-white region, 
  are both foreground stars.
 In panel b, the bright FUV emission at the location of the blue overdensity is emphasized
  and its alignment with the stellar stream. 
 In panel c, the the NGC\,5387 disk is shown in dark grey, the stream in light grey, 
  the blue overdensity in light blue, and the two foreground stars in yellow and white.
 }
 \end{figure*}

\begin{figure*}[tb]
\begin{tabular*}{1.\textwidth}{lcr}
 \includegraphics[width=.28\textwidth]{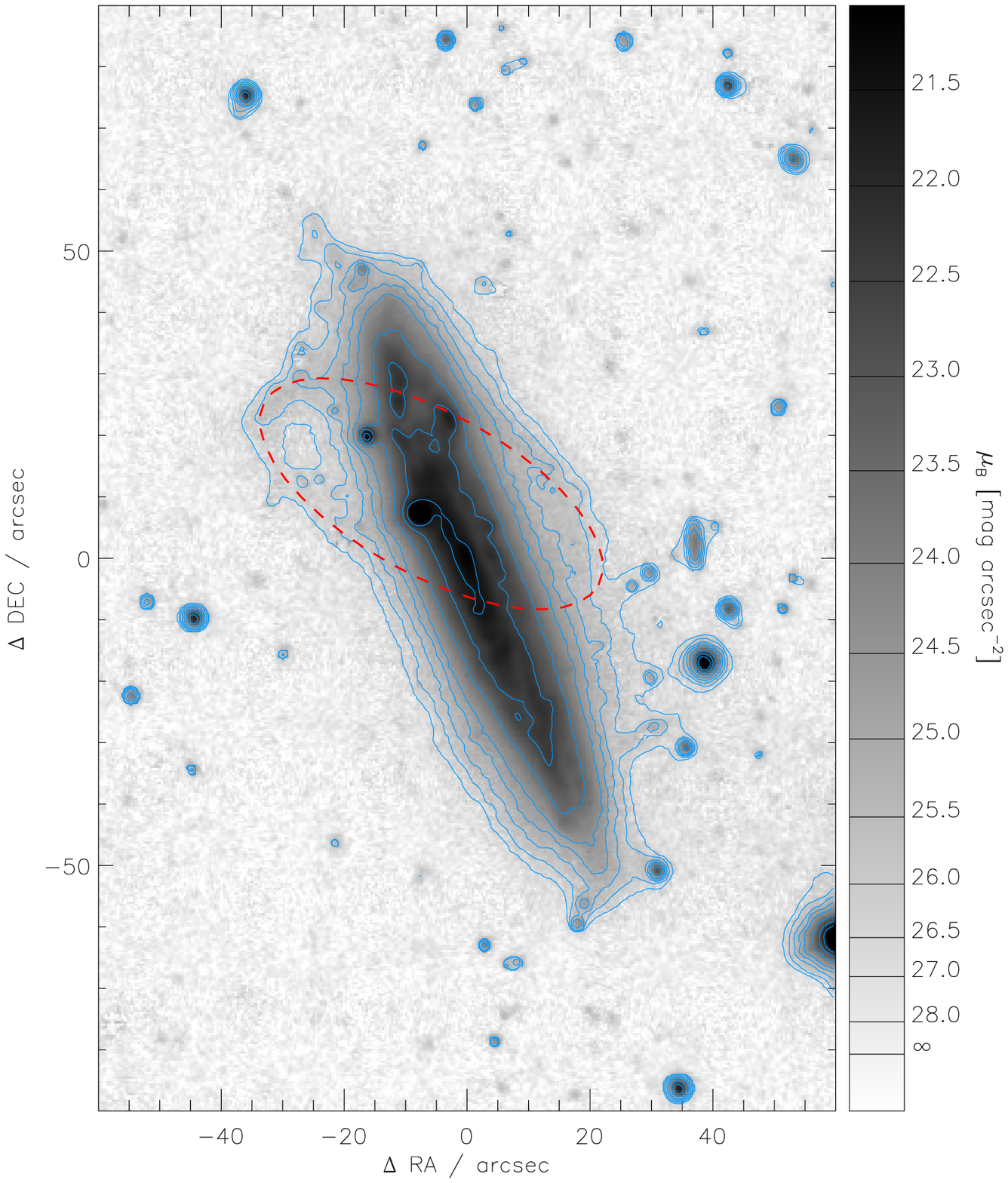} &
 \includegraphics[width=.28\textwidth]{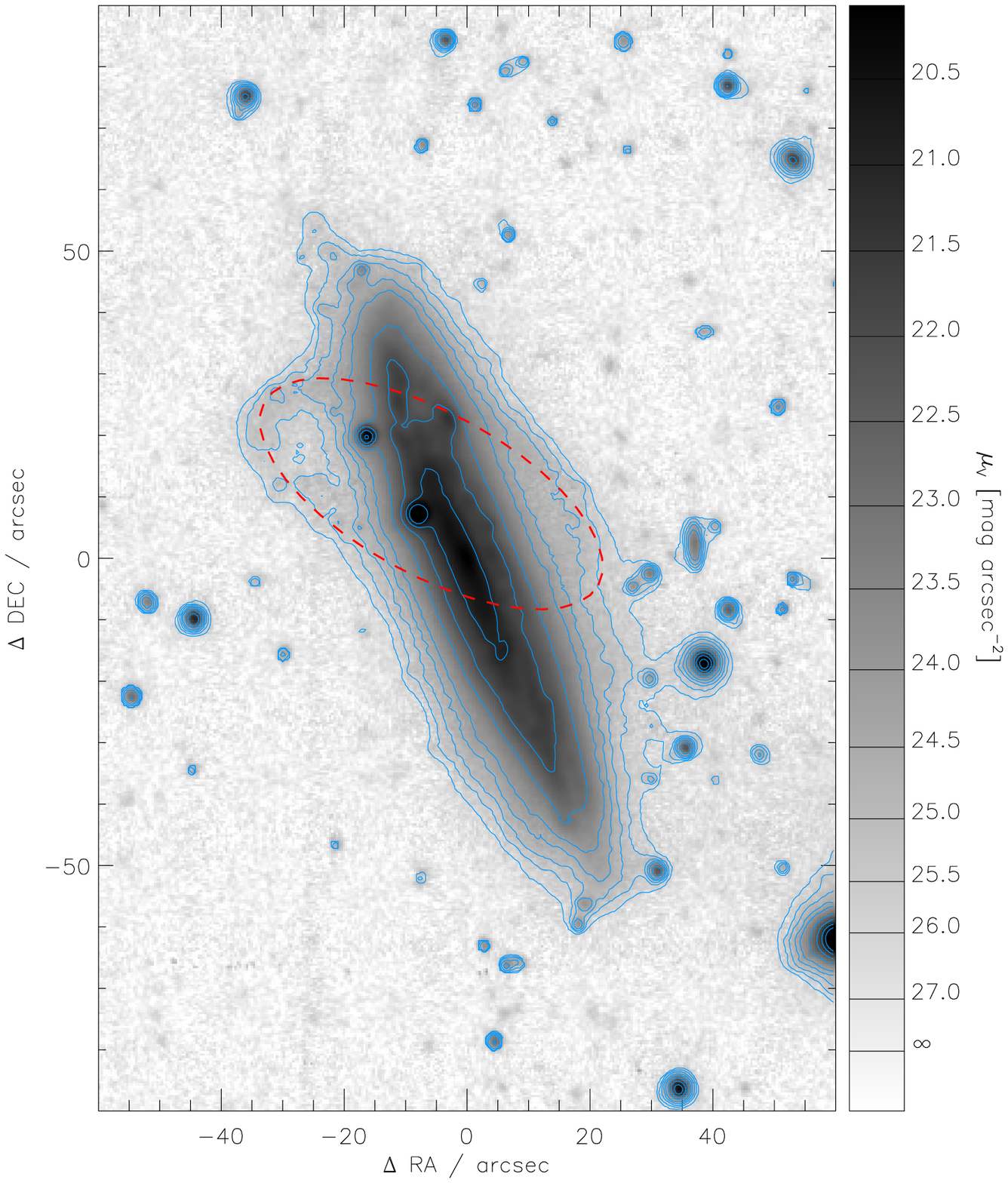} &
 \includegraphics[width=.28\textwidth]{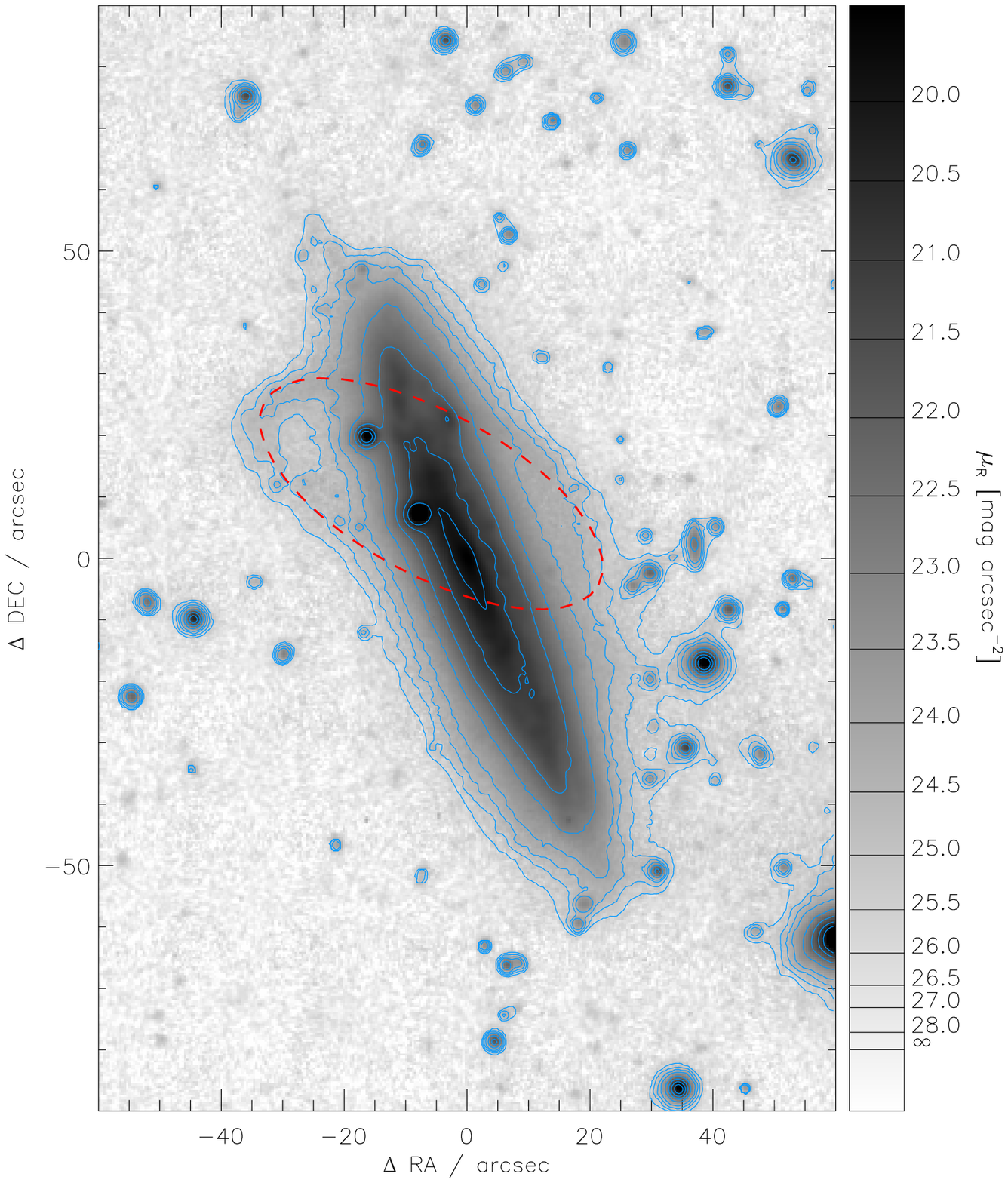} \\
\end{tabular*}
\caption{ \label{fig:surfdens} 
 VATT $B$ (left), $V$ (center) and $R$ (right) imaging of NGC\,5387. 
 The images are astrometrically aligned and calibrated for surface brightness. 
 The grey scale-bar on the right displays the surface brightness with asinh stretch
  as indicated in the sidebar. 
 Contours (in light blue) are shown at the following surface brightness values, 
  in mag arcsec$^{-2}$: (26.5, 26.0, 25.5, 25.0, 24.5, 23.5, 22.5, 21.5), 
  (26.0, 25.5, 25.0, 24.5, 24.0, 23.0, 22.0, 21.0, 20.0), 
  and (25.5, 25.0, 24.5, 24.0, 23.0, 22.0, 21.0, 20.0), for $B$, $V$ and $R$ respectively. 
 An ellipse (red dashed), as described in the text, is overlaid on the stream to
  emphasize the stream and its alignment with the blue overdensity.
}
\end{figure*}

\section{Revealing the Accretion Event in NGC\,5387}\label{sec:betterdata}

NGC\,5387 was imaged in the SDSS footprint and this imaging suggested the 
 presence of a low surface brightness stellar stream.
In Section~\ref{sec:NGC5387}, we first characterize NGC\,5387 based on archival data.
To confirm the stream, deeper observations were obtained 
 with the Vatican Advanced Technology Telescope (VATT) and
 are described in Section~\ref{sec:imgdata}.
Analysis of the imaging includes,
 the surface photometry of the stream presented in Section~\ref{sec:imagingresults}
 and estimation of progenitor properties from the stream morphology in Section~\ref{sec:stream}.
In addition to the stream, we explore the photometric properties of an overdense
 blue region coincident with the projected intersection of the stream with the disk. 
In Section~\ref{sec:blob}, we explore the photometric properties of this region. 

\subsection{NGC\,5387}\label{sec:NGC5387}
 
NGC\,5387 is an edge-on ($i = 80^{\circ}$), Sbc type galaxy \citep{RC3} with a warped disk \citep{warped_disk}.
Assuming the Tully-Fisher derived distance \citep{distance1}, the galaxy has a major axis diameter of 41 (48) kpc in $B$ ($K_s$)
 imaging \citep{RC3,LGA2MASS}.
Using the maximum rotation velocity, $V_{c} = 162.9 \pm 3.0$ km s$^{-1}$, 
 measured from the neutral hydrogen line profile,
 the total mass of NGC\,5387 is $1.08 \times 10^{11} M_{\odot}$ \citep{h1profile}.
The total HI flux, $9.79$ Jy km s$^{-1}$, yields a total HI mass of $1.46 \times 10^{10} M_{\odot}$ \citep{h1profile}.
Using the SDSS MPA-JHU Value Added Catalogs\footnote{http://www.mpa-garching.mpg.de/SDSS/DR7/},
 the total stellar mass is $2.75 \times 10^{10} M_{\odot}$
 \citep{sdss_dr7,Brinchmann2004}.
Thus, the neutral gas fraction is $M_{HI}/M_{*}$ = 0.53.

In the detailed spectroscopic diagnostics of \citet{Kauffmann2003}, NGC\,5387 is found to have
 no LINER or AGN component and have a central star formation rate (SFR, hereafter), SFR = 0.22 M$_{\odot}$ yr$^{-1}$.
The global SFR for NGC\,5387 can be estimated from its IRAS 100$\mu$ flux, 
 f$_{100\mu}$=1.22 Jy $\pm$ 0.146 Jy \citep{iras_cat}, to be SFR(FIR) = 2.5 M$_{\odot}$ yr$^{-1}$ using 
 the conversion for galaxies later than Sb \citep{fir_sfr}. 
Overall NGC\,5387 is of similar physical size as the Milky Way (MW), but is
 nearly an order of magnitude smaller in total mass \citep{MWmass2013}.
 
While NGC\,5387 has a typical $M_{HI}/M_{*}$ ratio for its stellar mass,
 its HI gas fraction is much larger than that of the MW \citep{blanton_review}.
Likewise, the oxygen nebular abundance from the SDSS MPA-JHU VAC, 12 + log(O/H) $= 9.05$,
 while higher than average, is not an outlier on the mass-metallicity trend established by \citet{Tremonti2004}.
Its total star formation rate is not 
 atypical for galaxies of its type \citep{Kauffmann2003}.
In its global physical properties, NGC\,5387 is on the whole smaller
 than the MW, but is ``normal'' compared to galaxies of similar mass.
Thus, we proceed to consider it a MW analogue galaxy.

\subsection{New Imaging Data}\label{sec:imgdata}
Deep optical imaging was acquired for NGC\,5387 at the Vatican Advanced Technology Telescope (VATT)
 with the VATT4k imager from UT 2012 March 21-28.
The VATT4k imager has a field of view of $12'$ $\times$ $12'$,
 and given its major axis diameter of $90''$, NGC\,5387 is well contained in a single pointing of the instrument. 
The plate scale is $0.37''$ per pixel at the 2$\times$2 binning mode used for these observations.
The imaging data were taken in the $B$, $V$, and $R$ filters with median seeing of $1.0''$. 
Individual exposures were $300$s or $450$s in length, with  
total exposure times of $8700$s, $9600$s, $9300$s in $B$, $V$, $R$ respectively.
The data were dithered randomly $\sim$1$'$ between exposures.  
``Off'' galaxy, or blank sky, exposures were acquired at least one field of view distant 
 from the target, but containing no significant extended objects or saturated stars.
Individual exposures were of the same length as individual ``on'' galaxy exposures
 (either $300$s or $450$s), but the total time exposed ``off'' galaxy 
 was roughly 25\% of the time spent ``on'' galaxy.

These imaging data were processed using image processing routines in the IRAF package MSCRED
 following the procedures outlined in the NOAO Wide-Field Survey (NWFS)
 project\footnote{\url{http://www.noao.edu/noao/noaodeep/ReductionOpt/frames.html}}.
The proceedures outlined for the NWFS were adapted for the VATT4k imager,  
 and only deviations from the standard process are described here.

The most significant deviation from the NWFS reduction is the absence of 
 a domeflat calibration, because there is no suitable dome flat screen at the VATT. 
Thus, inital flat-fielding was completed using high $S/N$ twilight exposures 
 collected throughout the observing run in all filters.
Object masks were created for each of the images and those images whose objects
 were well masked were median combined (typically $20-30$ per filter) to create a preliminary flat field for each filter.
After application of the primary flat field,  a super sky flat was created
 by applying the flatfield creation proceedure to the science and ``off'' galaxy exposures. 
After application of the appropriate sky flats,  data from the two amplifiers were merged to create a single FITS image for each frame, 
 to which a bad pixel mask was applied to remove known image defects. 
All images for a filter were registered with IMALIGN and then co-added using IMCOMBINE in IRAF.
The CCREJECT algorithm was applied during the stacking to remove cosmic rays and, given the large number of images
 in each filter, was highly successful. 

The images were astrometrically registered to SDSS by cross correlating the point sources detected on the VATT images
 against those in the SDSS-DR9 photometric catalogs \citep{dr9_cat}.
The photometric calibration is also performed by comparison
 to the point sources in the SDSS DR9 photometric catalog, after
 discarding those sources that are saturated or approached the
 non-linearity regime in the VATT images.
The resulting zero points are corrected to include 
 a term for the foreground Galactic extinction ($E(B-V)$=0.03).

To estimate the depth of the VATT images, we compute the
 surface brightness (SB) corresponding to the pixel-to-pixel root mean square variation (r.m.s.)
 to obtain 27.4, 27.0 and 26.7 mag arcsec$^{-2}$ in $B$, $V$, $R$ respectively. 
The pixel-to-pixel r.m.s. is a good measure of the SB sensitivity for a single pixel in the image. 
However, because we are interested in the SB averaged over many pixels, 
 the true limit to our measurements is not represented by the photon noise in a 1-pixel
 scale; instead our measurements are limited by any large scale background fluctuations. 
Hence, we have estimated the limiting SB as the r.m.s. of the median SB determined
 in a series of $\sim$10$'' \times \sim$10$''$ boxes of ``empty sky'' in fields around NGC\,5387.
The corresponding values in mag arcsec$^{-2}$ are 29.6
  29.6 and 28.3 for the $B$, $V$, $R$ images respectively.

\subsection{A Better View of the NGC\,5387 System}\label{sec:imagingresults}

Figure~\ref{fig:prettyimage} presents composite images of the NGC\,5387 system.
In Figure~\ref{fig:prettyimage}a, the VATT $R$ band image is shown in greyscale 
 with an SDSS color image of the NGC\,5387 disk inset.
In Figure~\ref{fig:prettyimage}b, the VATT $R$ band and SDSS image are 
 combined with GALEX FUV imaging of the region from the Deep Imaging Survey \citep[DIS, ][]{galex}.
Two key features of NGC\,5387 can be identified in Figure~\ref{fig:prettyimage}: 
 (i) a smooth stellar stream of ``great circle'' morphology \citep{streamsurveypaper}, 
 and (ii) a blue overdensity at the intersection of the stream and the disk.
Both of these regions are indicated in Figure~\ref{fig:prettyimage}c,
 where the NGC\,5387 disk is in dark grey, the stream in light grey
 and the blue overdensity as the elliptical blue region.
In addition to these features, 
 we note that the very circular blue and orange regions on the disk are both foreground stars 
 from their spectral energy distribution (SED) in their SDSS photometry.
These two stars are indicated in Figure~\ref{fig:prettyimage}c as the 
 white and yellow circles, respectively.
We will further describe the stream in Section \ref{sec:stream} and the blue overdensity in Section \ref{sec:blob}. 

A surface photometry analysis is performed on the VATT images for the NGC\,5587 disk. 
Total integrated magnitudes for the NGC\,5387 disk are measured
 by integrating the background-subtracted flux (Figure 2) over an
 elliptical aperture with semi-major diameter 90$''$, 
 corresponding to the major axis diameter \citep{RC3},
 and semi-minor diameter of 25$''$, corresponding to its minor axis diameter \citep{RC3}.
This ellipse corresponds to
 an $R$-band isophotal level of 27 mag arcsec$^{-2}$, i.e., $\sim3\sigma$ 
 above the large-scale background fluctuations. 
The total integrated apparent magnitudes for 
 NGC\,5387 are 14.46, 13.67 and 13.10 in $B$, $V$, $R$, respectively. 
Using the relations of \citet{ZCR09}, 
 the integrated colors and assuming a distance modulus, ($M-m$) = 34.5, 
 we obtain an average mass estimate of $3.13\times10^{10} {M_\odot}$ 
 with a 40\% variation from the choice of color and reference magnitude 
 used in the adopted $M/L$-color relation. 
The surface photometry for NGC\,5387 is summarized in Table 1.

\begin{figure*}
 \begin{tabular*}{1.\textwidth}{lcr}
  \includegraphics[width=.28\textwidth]{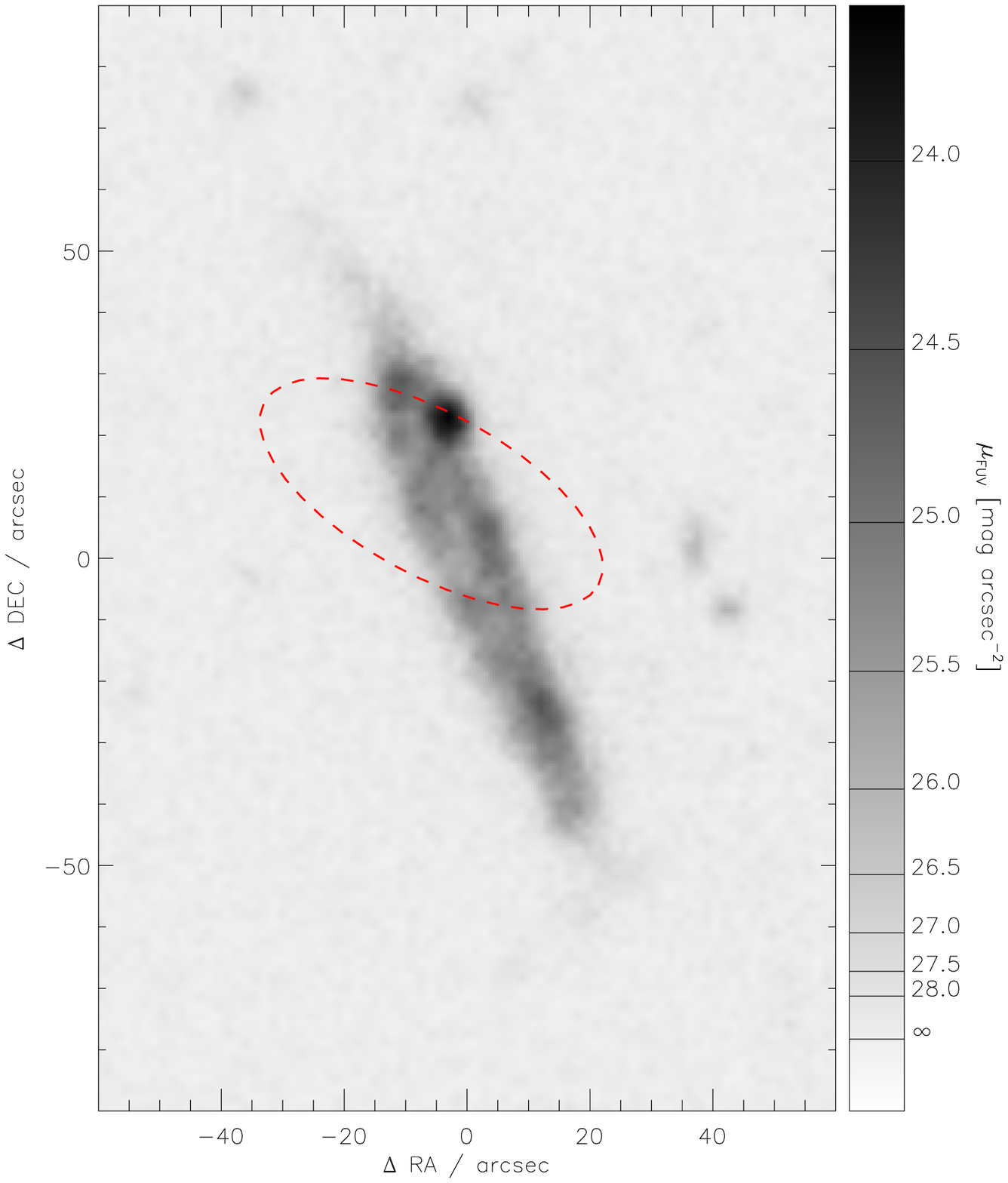} &
  \includegraphics[width=.28\textwidth]{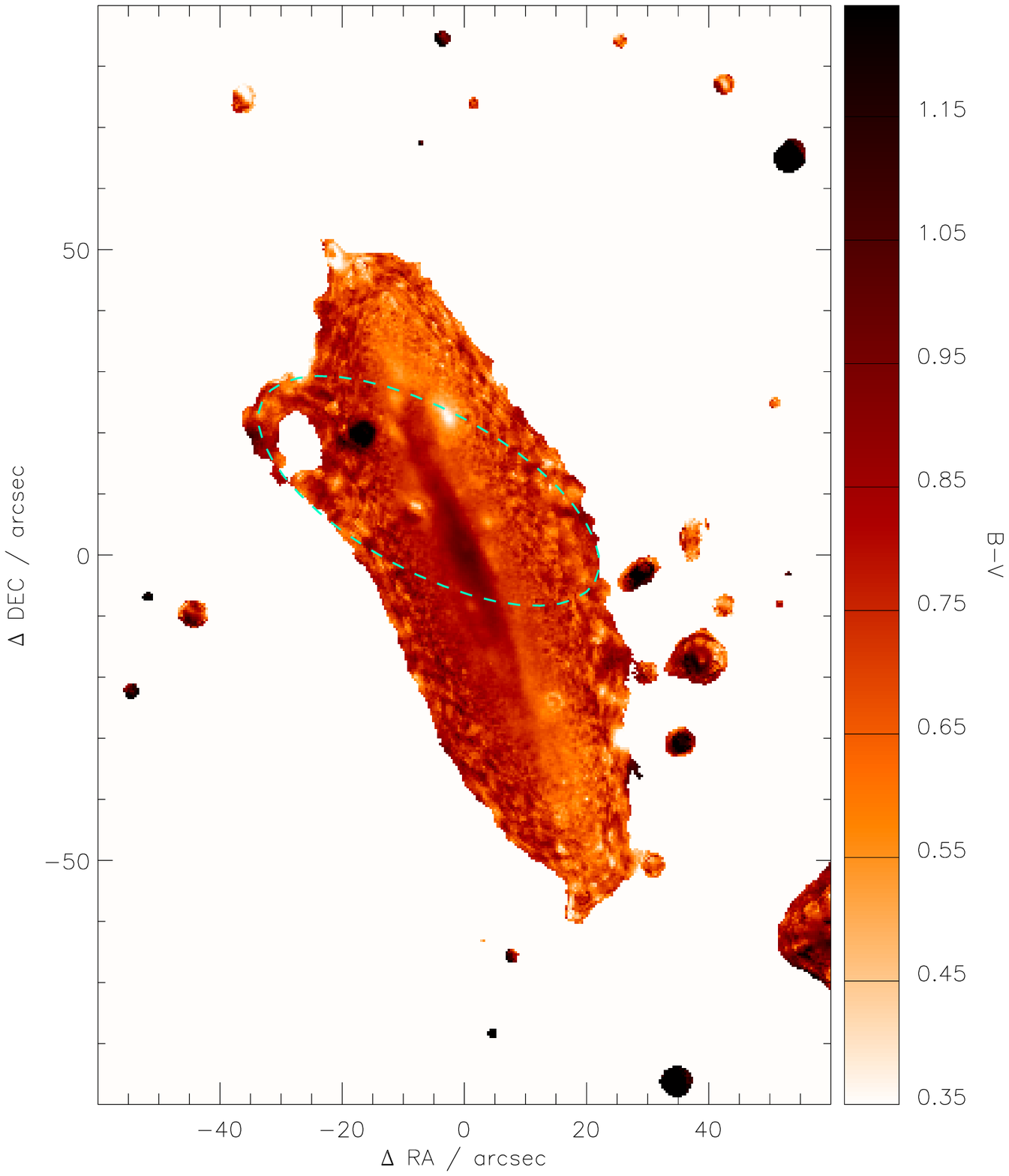} &
  \includegraphics[width=.28\textwidth]{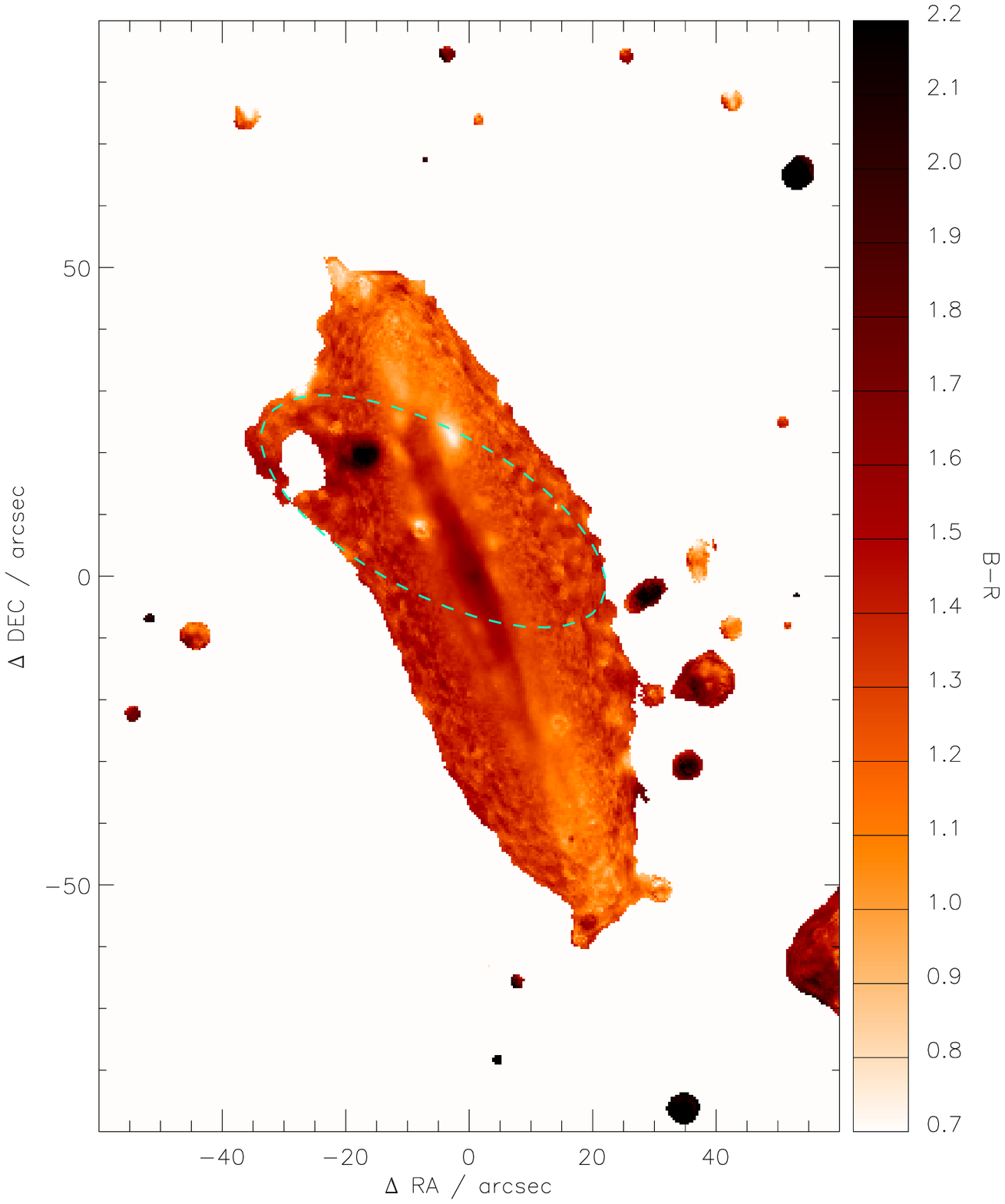} \\
 \end{tabular*}
 \caption{ GALEX FUV (left), $B-V$ (center) and $B-R$ (left) maps of NGC\,5387.
   As in Figure \ref{fig:surfdens}, the maps are astrometrically registered and 
    an ellipse is over-plotted to indicate the position of the stellar stream 
   (red dashed on the left and green dashed in the center and right panels).
  The blue overdensity as described in the text is very bright in the FUV image (left) and  
   has $B-V$ and $B-R$ (center, right) colors remarkably dissimilar 
   from the mean colors of NGC\,5387.
  The blue overdensity is also different from the color of stream, 
   which is overall similar to those of the NGC\,5387 disk.}
\end{figure*}

\subsection{The Stellar Steam of NGC\,5387}\label{sec:stream}

The greyscale representation of the $R$ band imaging of Figure 1 clearly
 shows the presence of a stellar stream extending to the east of the
 Northern half of the NGC\,5387 disk, and extending through the disk to the west. 

The morphology of a stream is difficult to often interpret 
 due to the finite surface brightness limit of the observations.
More specifically, we are biased toward seeing only the brightest segments of the stream, which 
 typically correspond to the ``youngest'' segment --- that is 
 those sections  most recently unbound from the progenitor \citep{kvj_2008,streamsurveypaper}.
The surface brightness limit combined with line-of-sight inclination effects
 often mislead the interpretation of the stream morphology, more specifically the
 determination of its full angular extent, number of wraps, and its total mass.
In the case of the stream in NGC\,5387, we see a very clear narrow portion to the northeast of the disk
 and fainter segments on the western side of the NGC\,5387 disk.
We only identify a single wrap of the stream that can be traced for a full 360$^{\circ}$.
Given that identified segments are measured at low signal-to-noise, 
 we cannot rule out the existence older wraps at even lower surface brightness.

The stream in NGC\,5387 is most consistent with a ``great circle'' morphology,
 which implies the structure was formed from a satellite in an approximately 
 circular orbit \citep{kvj_2008,streamsurveypaper}. 
The stream appears to be elliptical in projection and can be described by an ellipse with semi-major
 axis of 82.7$''$ (32 kpc), an axial ratio of 0.42, 
 and a position angle 61.3$^{\circ}$ (measured from North to East).
This ellipse is indicated in Figures 2 and 3. 
We will now determine properites of the stream and its 
 progenitor inferred from its ``great circle'' morphology (Section 2.4.1) and measured with
 surface photometry (Section 2.4.2).
 
\subsubsection{Morphology of the Stellar Stream} \label{sec:streammorph}

From the morphology of the stream (see Figure 1), we can derive an 
 estimate of the progenitor mass ($m_{sat}$) and ``age'' of the stream ($t_{str}$) 
 following the analytical relations derived by \citet{kvj_2001}.
For the purposes of this paper the age of the stream, $t_{str}$, refers
 to the time since the accretion of the progenitor, not the 
 age of the stellar populations comprising the stream.

Assuming a logarithmic potential for the host and $M_{NGC\,5387} >> m_{sat}$, 
 then the mass, $m_{sat}$, of the stream progenitor, is estimated by:
 \begin{equation}\label{eq:streammass}
  m_{sat} \sim \left(\frac{w}{R}\right)^3 \left(\frac{R_{peri}}{10~kpc}\right) \left(\frac{v_{circ}}{200~km/s}\right)^{2} 10^{11} M_{\odot} 
 \end{equation}
where $w$ is the width of the stream at $R$, $R_{peri}$ is the radius at pericenter,
 and $v_{circ}$ is the circular rotation velocity of the parent.
Likewise, the age $t_{str}$ of the streamer can be estimated as:
 \begin{equation}\label{eq:streamage}
t_{str} \sim 0.01 \Psi \left(\frac{R}{w}\right)  \left(\frac{R_{circ}}{10~kpc}\right) \left(\frac{200~km/s}{v_{circ}}\right) Gyr 
 \end{equation}
 where $\Psi$ is the angular length of the stream and $R_{circ}$ is the 
 radius of a circular orbit with the same energy as the true orbit.

We adopt the circular rotation velocity $v_{circ}=$162.9 $\pm$ 3 km s$^{-1}$ \citep{h1profile}.
$\Psi$ can be estimated visually from Figure~\ref{fig:prettyimage} to be $\ge 360^{\circ}$.
Both $w$ and $R$ can be evaluated at any point in the stream
 and we choose the region of the stream with the highest apparent surface brightness (to the N of the NGC\,5387 disk)
 such that $w$~=11$''$= 4.2 kpc and $R$=16 kpc. 
Given the ``great circle'' morphology, the progenitor of the stream was most likely on a circular or near-circular orbit,
 and we can approximate $R_{circ}$ = $R_{peri}$ = $R$ = 16 kpc (based on the ellipse in Figures 2 and 3).
Substituting these values into Equation~\ref{eq:streammass}, we estimate a total 
 progenitor mass of 2~$\times$~10$^{10}$ M$_{\odot}$,
 and into Equation~\ref{eq:streamage}, the formation age for the streamer of $t_{str}\sim$~400 Myr.

\subsubsection{Surface Photometry of the Stellar Stream}\label{sec:streamsurfphot}

In Figure 2, we compare the stellar stream of NGC\,5387 as observed in
 our three VATT filters, $B$, $V$, $R$. 
In all three optical filters the stream is detected and appears to have similar overall morphology.
Overall the stream appears to be a single structure
 with mean surface brightness of 25.4 mag arcsec$^{-2}$ in $R$. 
Though, we note that there are significant density fluctuations 
 and the stream reaches 24.6 mag arcsec$^{-2}$ at its maximum surface brightness. 
By assuming a typical width of 11$''$ (5.1 kpc), the geometrical parameters given in Section 2.4 and an
 approximately uniform surface brightness over the entire elliptical
 path (which cannot be verified in the overlap region with the
 main galaxy), we can estimate a total magnitude for the stream of $\approx17.5$ mag in $R$.

To investigate the physical properties of the stream, we
 compute color maps from the VATT images. 
To do this, we enhance the original signal-to-noise ratio (SNR) in the individual
 pixels by performing the image adaptive smoothing introduced by
 \citet{ZCR09} using the {\sc adaptsmooth} code of \citep{adaptsmooth}. 
{\sc adaptsmooth} performs a median filtering of the
 images using a circular top-hat kernel, whose size is adapted as a
function of position to reach a given minimum SNR. 
 A first pass of {\sc adaptsmooth} is made in each band independently, to determine the
 ``mask'' of smoothing kernel sizes as a function of the position that
 provides a minimum SNR of 20. 
In the second pass the smoothing kernel
 size at a given pixel is re-determined as the maximum size over the
 three bands, so that the three images are smoothed consistently while
 ensuring the minimum SNR of 20 in all pixels.  
The $B-V$ and $B-R$ color maps are shown in the second and third panel of Fig. 3, respectively. 
From these maps for the stream we can estimate a
 characteristic $(B-V)_{str}$ between 0.6 and 0.9, and
 $(B-R)_{str}$ between 1.2 and 1.7. 
These colors are broadly consistent with the average colors of NGC\,5387, i.e., $(B-V)_{gal}=0.79$ and
 $(B-R)_{gal}=1.36$, thus implying (approximately) similar stellar populations.
 This observation justifies assuming the same $M/L$ ratio for the stream
  as for the galaxy: the stellar mass ratio between the streamer and
  NGC\,5387 is thus given by their luminosity ratio, i.e.~$\sim$1:50, 
  implying a mass for the streamer of $\sim$6 $\times$ 10$^{8}$ ${M_\odot}$.
These photometric properties are summarized in Table 1.

\subsubsection{Summary of Stellar Stream Properties}

The morphology of the stream implies that the progenitor was accreted within the last Gyr ($t_{acr}$),
 an age that is broadly consistent with its relatively small $R_{peri}$. 
There are no distinguishable overdensities within the angular extent of the stream,
 and, therefore, we do not identify a satellite remnant within the stream itself.
In some known stellar streams, the progenitor is either nearly unbound or 
 can be very elongated in the plane of the sky,
 as is implied by detailed simulations for the stream in NGC\,5907 \citep{dmd2008}.
Given the median seeing ($\sim 1''$) and the pixel scale, 0.38$''$ per pixel, 
 a highly distorted progenitor would be difficult to identify in our observations.
Furthermore, the lack of progenitor is not uncommon, as 
 most of the streams simulated in \citet{kvj_2008} had no identifiable progenitor.
Comparing the mass of the progenitor from Equation 2 with the
 mass estimate of the stream itself (Section\ref{sec:streamsurfphot}),
 we estimate that the stream contains $\sim$3\% of the total mass of the progenitor.

The great circle morphology of the NGC\,5387 stream is similar to those identified 
 around M\,63 \citep{chonis2011}, NGC\,5907 \citep{dmd2008}, and NGC\,4013 \citep{dmd2009},
 but with a smaller apocentric radius and implied formation timescale.
The inferred progenitor mass is comparable to that predicted for 
 Sagittarius around the MW \citep{law2005}, though the Sagittarius debris have multiple wraps
 and are thought to have formed over several Gyrs \citep{2massSgr}, in contrast to 
 the single wrap and short formation timescale for the NGC\,5387 stream.
Only one wrap of the NGC\,5387 stream is detected in our observations, 
 but we cannot rule out that there are older wraps of the stream,
 as was discovered for NGC\,5907 when deeper observations were 
 obtained \citep{shang1998,zheng1999,dmd2008}.

\subsection{Properties of the Blue Overdensity}\label{sec:blob}

In Figure~\ref{fig:prettyimage}a,
 there is a blue region offset 23.6$''$ radially from the center of NGC\,5387, 
 corresponding to a projected galactocentric radius of 22.7$''$. 
The region is roughly circular with a radius of 2.2$''$.
Given the median seeing of 1$''$, we are not able to resolve any sub-components to the blue overdensity
 and we treat it as a single region.

Using a circular aperture of 2.2$''$ the total $B$, $V$, $R$ absolute magnitudes 
 are -$14.42$, -$14.72$, -$14.89$, assuming 
 a distance of 79 Mpc and accounting for the MW reddening of $E(B-V)_{MW} = 0.03$ with a standard
 \citet{CCM} extinction curve for $R_{V} = 3.1$.
The blue overdensity has a ($B-V$) = 0.30, implying $\Delta(B-V)$ from the stellar stream of $\sim$0.30.
If the blue overdensity is part of the stream, then this is the first known stream with 
 such an extreme color gradient ($\Delta$(B-V)$\sim$ 0.3-0.6).  

In Figure 3, we compare our optical imaging to that of the GALEX Deep Imaging Survey \citep{galex}
 to the $B-V$ and $B-R$ color maps.
In Figure 3, it becomes clear not only that the region is much bluer than the disk of NGC\,5387 and the stream, 
 but also that the region contributes a significant portion of the FUV flux from NGC\,5387.
The total integrated FUV (NUV) magnitude of NGC\,5387 is m$_{FUV}$ = 17.44 (m$_{NUV}$ = 16.83)
\citep[this estimate includes the blue overdensity][]{lemonias2011}.
In comparison, the total FUV (NUV) magnitude for the blue overdensity is m$_{FUV}$ = 18.49 (m$_{NUV}$ = 18.21).
From the apparent magnitudes, the blue overdensity contributes 38\% (28\%) of the FUV (NUV) 
 flux of NGC\,5387. 
The blue overdensity is most likely a star forming region, but from the imaging alone it is 
 impossible to determine if it is related to NGC\,5387 or an overlapping foreground object. 


 \begin{figure*}[tb]
 \plotone{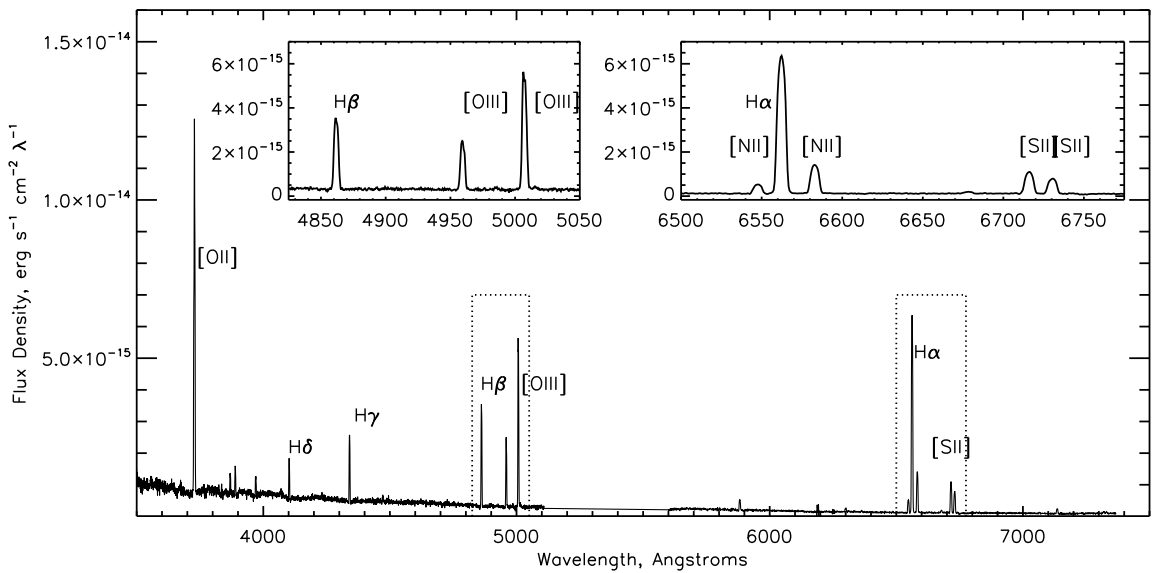}
 \caption{\label{fig:spec} Rest frame wavelength LBT+MODS spectrum for the blue overdensity in the outer disk of NGC\,5387, 
   with insets of the H$\alpha$ and H$\beta$ spectral windows.
  The spectrum exhibits nebular emission lines consistent with an HII region.
 }
 \end{figure*}

\section{The Bright Star Forming Region in NGC\,5387}\label{sec:spec}
 
 To better characterize the blue overdensity and its potential connection 
 with the stellar stream, we obtained optical spectroscopy for the region. 
 We will first describe the spectroscopic observations in Section~\ref{sec:modsdata} and
  our measurement proceedures in Section~\ref{sec:redshift}.
 In Section~\ref{sec:specresult} we describe initial measurements from the spectroscopy.
 We estimate the physical parameters of the blue overdensity directly from the 
  spectra in Section~\ref{sec:sbprop} and from comparison to Starburst 99 models in Section~\ref{sec:sb99prop}.
 Lastly, we explore the mode of star formation in this region in Section~\ref{sec:sfmode}.

\subsection{Observations \& Image Processing}\label{sec:modsdata}
 Spectroscopy was obtained for NGC\,5387 using the dual channel MODS1 spectrograph on the 
  Large Binocular Telescope (LBT) \citep{modsref} on UT 2012 June 14.
 The spectrograph was configured with the $R$ = $2000$ resolution grating and a $1.0''$ slitwidth,
  yielding a spectral range of  $3200\AA$ to $10000\AA$
  with mean resolution of $0.5\AA$, $0.8\AA$ per pixel for the central wavelength of the
  blue and red channels, respectively.
 Two pointings were obtained, one along the major axis of NGC\,5387 (position angle $= 22^{\circ}$)
  with central coordinates 
 ($\alpha_{J2000}, \delta_{J2000}) = $(13$^h$58$^m$24.8$^s$, +06$^{\circ}$04$'$17$''$)
  and one parallel to the major axis offset by $5.8''$ to intersect the blue overdensity in the disk.
 Total exposure times were $3 \times 240s$ and $3 \times 600s$ for the two pointings, respectively.

 Image processing was completed using the MODS1 python
  packages\footnote{http://www.astronomy.ohio-state.edu/MODS/Software/modsCCDRed/}
  following the standard prescription for MODS1 spectroscopic data.
 The one-dimensional spectra were extracted for a $2''$ region of the NGC\,5387 nucleus
  and the blue region in the APEXTRACT package of IRAF with independent extractions for the blue and red channels.
 Wavelength calibrations were taken with the $0.6''$ slit and a small zero-point offset
  was calculated between the wavelength calibration spectrum and the science data using the night sky lines
 (the standard proceedure for this instrument).
 Observations of the spectro-photometric standards Wolf 1346 and Hz 44 were obtained on the same night of observation
  and were used to spectro-photometrically correct the spectra for extinction 
  using calibration data available in IRAF.
 Sensitivity functions were calculated independently for the two stars,
  and were averaged before application to the science spectra.

\subsection{Spectral Measurements}\label{sec:redshift}

 The line-of-sight velocities of the galaxy from the MODS1 spectra were computed using the FXCOR routine in IRAF.
 An emission line template was created for emission lines in the
  $H\beta$ and $H\alpha$ regions smoothed by the average instrumental broadening of $3.36\AA$, 
 which was estimated from lines in the wavelength calibration spectrum and the night sky lines in the science spectrum.
 The barycentric correction for the median time of each set of exposures was applied to this template.

 We estimate the heliocentric velocity of NGC\,5387 to be $v_{helio} = 5226 \pm 3$ km s$^{-1}$, well in agreement
   with the systemic velocity derived from its HI profile, $5216$ km s$^{-1}$ \citep{h1profile}
   and that measured from the SDSS spectrum $5226$ km s$^{-1}$ \citep{sdss_dr7}.
 Using the same proceedure, the heliocentric velocity of the blue overdensity was measured to be $5331 \pm 2$ km s$^{-1}$.   

 \begin{figure*}[tb]
\plotone{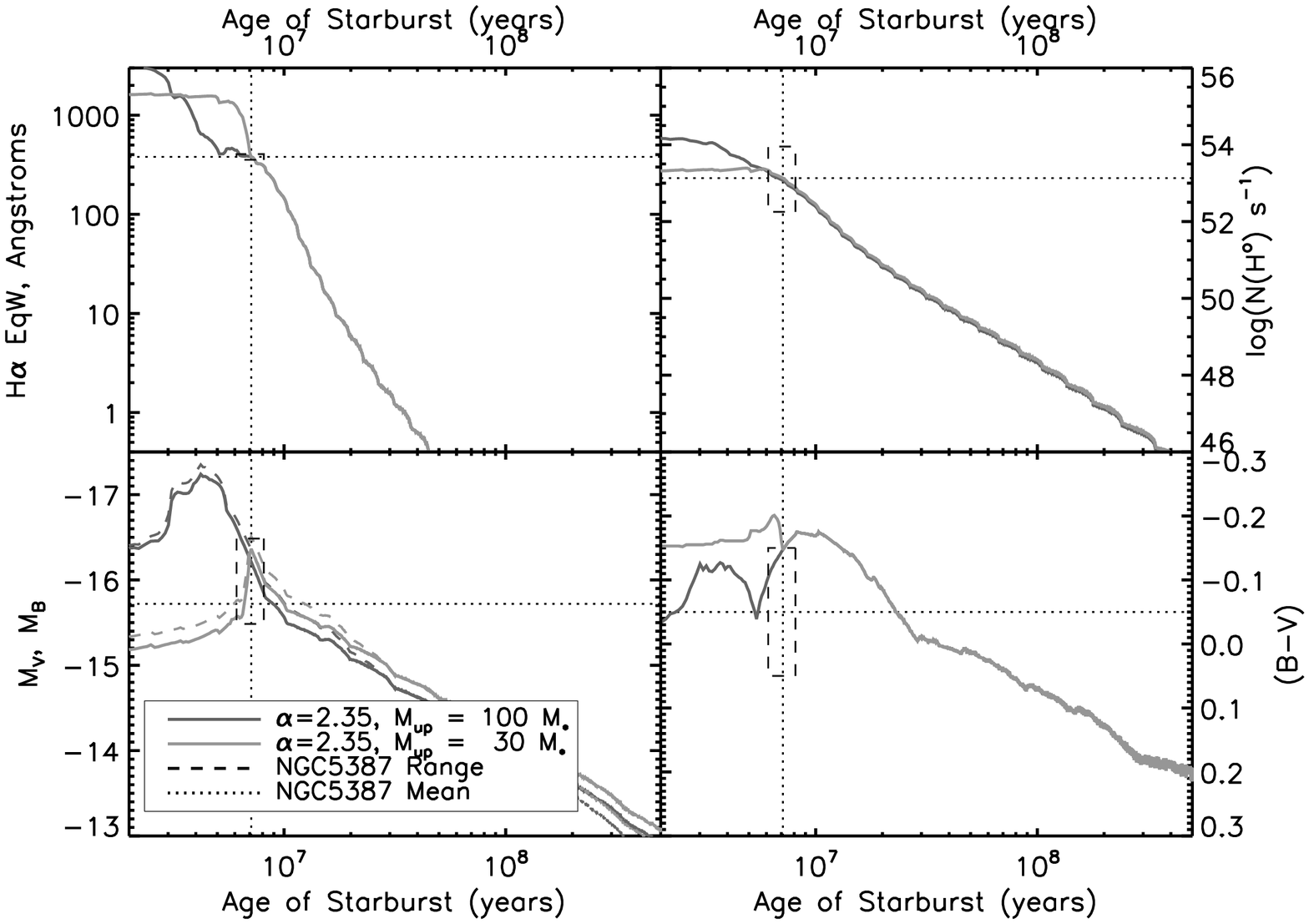}
 \caption{\label{fig:sb99} Comparison of the spectral properties of the NGC5386 blue overdensity to the fiducial Starburst 99 model,
  with a Salpeter IMF ($\alpha$ = 2.35) and with upper mass cutoffs of 
  100 M$_{\odot}$ (dark grey) and 30 M$_{\odot}$ (grey). 
 In the top panels, the H$\alpha$ equivalent width ($EqW_{H\alpha}$, left) 
  and the number of ionizing photons per second ($N(H^{\circ}) s^{-1}$, right)
  are used to estimate the age of the star forming region, resulting in a mean value of 8 $\times$ 10$^{6}$ years.
 In the bottom left panel, the age of the star forming region is compared to the total V (B) band magnitude, $M_V$ ($M_B$)
  as the solid (dashed) thick lines.
 In the bottom right panel, the age of the star forming region is compared to the model $(B-V)_o$ colors
  to confirm the internal reddening derived from the Balmer decrement method.
 In each panel the the ``mean'' value from comparison is shown as a dotted line, 
  and the permitted ``range'' of values permitted from the uncertainty in each of the 
  the measurements are shown as a dashed box. }
 \end{figure*}
 

 Before estimation of emission line fluxes, we corrected the MODS1 spectra for 
  the local Galactic extinction, $E(B-V) = 0.03$ \citep{SFD_redo, SFD},
  assuming $R_{V} = 3.1$ and applying the extinction law of \citet{CCM}.
 
 Equivalent widths for H$\alpha$ and H$\beta$ were measured using the interactive IRAF task SPLOT. 
 Generally, standard line profiles available in SPLOT were poor fits to the line shapes.
 Thus, emission line fluxes were calculated independently of SPLOT by integrating across the line profile 
  and subtracting the best fit continuum defined individually around each emission line.
 The median continuum level was 3.85$\times$10$^{-15}$ and 9.75$\times$10$^{-16}$ 
  erg s$^{-1}$ cm$^{-2}$ $\AA^{-1}$ for the blue and red channels respectively. 
 There was no identifiable stellar absorption at any of the emission lines,
  (perhaps because we did not integrate sufficiently long to detect the underlying stellar spectral features),
  but the contributions from the stellar absorption to the measurements required for our work would be small.
 Thus, we opted not to fit a stellar absorption component to the spectra. 

 Flux errors were estimated by adapting the simple formulation of \citet[Equation 2]{berg2012}:
 \begin{equation}
  \sigma_{\lambda} \sim \sqrt{(2 \times \sqrt{N} \times F_{rms})^2 + (0.02 \times F_{\lambda})^2}
 \end{equation}
  where $N$ is the number pixels in the integration, $F_{rms}$ is the median noise in the continuum, 
  0.02 is the assumed error (2\%) in the summation relative to the total measured flux ($F_{\lambda}$).
 Errors on derived values are propagated using $\sigma_{\lambda}$ as the flux error.

 The SDSS imaging was used to design the MODS1 observational parameters. 
 Given the SDSS pixel scale, $0.39''$, and low signal-to-noise detection of the blue overdensity, 
  the offset between the NGC\,5387 major axis and the overdensity was imperfect.
 Furthermore, the slit was only $1''$ in angular width and therefore the 
  spectroscopic observations did not fully contain the blue overdensity. 
 A correction was computed using FUNTOOLS\footnote{https://www.cfa.harvard.edu/$\sim$john/funtools/},
  which uses the DS9 region style syntax to perform flux measurements with custom apertures.
 Using a circular aperture of $2.2''$, we first calculated the total flux in the blue overdensity,
  used a nearby region in the disk to estimate the background contributions from NGC\,5387.
 Then, using the MODS1 observing parameters, MODS1 target acquisition images 
  (images taken to align the slit on target), and the spectral extraction parameters in APEXTRACT, 
  the area covered by the slit, the area extracted in the slit, 
  and the area subtracted as a background were determined.
 The fluxes then were corrected using the ratio of the total flux of the star forming region
  and flux extracted from the spectra.
 The fluxes for both the main galaxy and the emission line region
  were converted to the corresponding luminosity assuming $D=79.3$ Mpc \citep{distance1,distance2}.

\subsection{Spectral Analysis Methods}\label{sec:specresult}

 The full spectrum for the star forming region is shown in Figure~\ref{fig:spec},
  with insets providing zooms into the H$\alpha$ and H$\beta$ regions.
 The velocity of the star forming region is offset $\delta$v = $-104$ km s$^{-1}$ from the NGC\,5387 systemic velocity.
 This offset is well within the NGC\,5387 HI profile full-width of $W$ = $350$ km s$^{-1}$\citep{h1profile}. 
 Thus, the blue overdensity is associated with the NGC\,5387 system.
 Assuming the blue overdensity is at the same distance as NGC\,5387, 
  the blue overdensity is approximately 850 pc in diameter
  and is located 9.1 kpc from the center of NGC\,5387 or 8.7 kpc projected onto the major axis
  with a disk height of 2.7 kpc.

 Having determined the blue overdensity is a star forming region,
  the assumed Milky Way extinction is not sufficient to account for the extinction in the region
  and it is necessary to estimate the internal reddening using the Balmer decrement method. 
 Following the prescription of \citet{calzetti2012} and taking 
  representative values from \citet{osterbrock}, we estimate 
  $H\beta/H\alpha = 4.15$, which when compared to the ideal value of 
  $H\beta/H\alpha = 2.87$ yields $E(B-V) = 0.35$ with an error of 0.08 (26\%) estimated
 by propagating uncertainties from the flux measurements through our calculations.   
 We then revise our photometry and fluxes using the extinction curve estimates
  for the Small Magellanic Cloud as detailed in \citet[Table 2]{Bianchi2011}.

 The measured reddening, $E(B-V) = 0.35$, implies an $A_V\sim1$, which is not
  abnormal for star forming regions \citep[for example see Table 2]{reines2008} ---
  though it is higher than might be anticipated given the strong FUV emission from the region (see Figure 3).
 Because the reddening is not abnormal, the star forming region is most likely 
  in front of the NGC5387 disk along the line-of-sight.

\subsection{Spectroscopic Properties of the HII Region}\label{sec:sbprop}

 From the fluxes of the nebular emission lines shown in Figure~\ref{fig:spec},
  we can directly calculate a number of properties of the region, 
  including the SFR, ionizing flux and chemical abundance.  
 All of these values are summarized in Table 2. 
 In the text we will apply our internal extinction correction of $E(B-V)$=0.38,
  but we also include values for only the Galactic extinction in Table 2 
  for comparison to systems without comparable measurements of internal extinction.

\smallskip
\smallskip
\noindent{\sc Star Formation Rate:}~ 
The SFR of the blue overdensity can be derived
 from the H$\alpha$ flux using the relationship of \citet{sfr_review} and \citet{osterbrock}: 
 \begin{equation}
  SFR(H\alpha) = 7.9 \times 10^{-42} L(H\alpha) 
\end{equation} 
Using our H$\alpha$ luminosity, 1.74 $\times$ 10$^{41}$ erg s$^{-1}$ with a 24\% uncertainty from flux and distance uncertainties, 
 we obtain SFR($H\alpha$) = 1.77 M$_{\odot}$ yr$^{-1}$ (log(SFR) = 0.25). 
An uncertainty of 0.43 M$_{\odot}$  yr$^{-1}$ in SFR($H\alpha$) is estimated by propagating 
 flux errors through the SFR calculation.

A second, independent estimate of the SFR can be calculated using 
 the GALEX FUV photometry in Table 2 \citep{galex}.
The reported FUV flux for the star forming region is 121.67 $\pm$ 0.83 $\mu$Jy 
 corrected only for Milky Way extinction. 
We estimate a background flux based on a similarly sized region in the disk 
 as 89.37 $\mu$Jy. 
Converted to luminosity, we obtain $L(FUV)$ = 1.95 $\times$ 10$^{28}$ erg s$^{-1}$.
Following the relation of \citet{sfr_review},
 \begin{equation} 
  SFR(FUV) = 1.4 \times 10^{-28} \times L(FUV)
 \end{equation}
 results in SFR(FUV) = 2.72 M$_{\odot}$ yr$^{-1}$ (log(SFR) = 0.44),
 a value in reasonable agreement with that from $L(H\alpha)$.
An uncertainty of 0.20 $M_{\odot}$ yr$^{-1}$ is estimated by propagating the 
 flux measurement error through the calculation.

\smallskip
\smallskip
\noindent{\sc Ionizing Flux:}~ 
To better understand the mode of star formation, 
  we also use $L(H\alpha)$ to estimate the total flux of ionizing photons, $N_{LyC}$ in the HII region. 
Following from \citet{Condon1992}, for a $10^{4}$ K gas, 
 \begin{equation}
  N_{LyC} \gtrapprox 7.87 \times 10^{11} \times L(H\alpha)
 \end{equation}
 we estimate a total ionizing flux of $N_{LyC} = 1.35 \times 10^{53}$ photons,
 with a 25\% error from the flux uncertainty.
Using the observations of \citet{Vacca1996} this is equivalent
  to $\sim 1500$ O$V7.5$ stars in our 850 pc diameter region.   

\smallskip
\smallskip 
\noindent{\sc Chemical Abundance:}~ 
The strong line calibration of \citet{pt2005} permits computation of the metallicity, 
 12 + log(O/H), from strong spectral features.
This method utilizes the spectral line ratios
 $R_{2}=[OII]_{\lambda3727+\lambda3729}/H\beta$, $R_{3}=[OIII]_{\lambda4959+\lambda5007}/H\beta$,
 and $R=[OIII]_{\lambda4363}/H\beta$
 to estimate the excitation parameter, $P=R_{3}/(R_{2}+R_{3})$, which is correlated to the oxygen metallicity.
Most notably, the calibration has upper and lower branches that diverge between 
 $X_{23}=\log (R_{2}+R_{3})$ of 0.7 and 1.0.
For our HII region we calculate, $P=0.34$ and $X_{23} = 0.83$, which place the region in the 
 transition zone between the two calibrations \citep[see Figure 12 of][]{pt2005}.
Thus, we opt to average $12 + \log(O/H)$ calculated on both branches, 
 for which we obtain $12 + \log(O/H) = 7.97$ for the lower branch 
 and $12 + \log(O/H) = 8.06$ for the upper branch. 
We adopt $12 + \log(O/H) = 8.03$ for the metallicity of our HII region,
 and adopt the 40\% error from the \citet{pt2005} calibration. 

\subsection{Properties Derived by Comparison to Starburst99}\label{sec:sb99prop}
 In addition to properties derived directly from flux measurements,
  star forming properties can be estimated in comparison to the fiducial 
  Starburst 99 models \citep[SB99, hereafter]{sb99}. 
 In Figure~\ref{fig:sb99} the model predictions from SB99 used in our analysis are 
  given for an instantaneous 10$^6$ M$_{\odot}$ starburst in a $Z = 0.001$ metallicity gas,
  with a Salpeter IMF ($\alpha$ = 2.35) and upper mass cutoffs of 30 M$_{\odot}$ (dark grey)
  and 100 (light grey) M$_{\odot}$.

\smallskip
\smallskip
\noindent{\sc Stellar Population Age:}~ 
 The age of the HII region can be estimated using the equivalent width
  of the H$\alpha$ emission line.
 We obtain the $EW_{H\alpha} = 380\AA$.
 In the top left panel of Figure~\ref{fig:sb99}, we compare our measurement (dotted lines)
  to the H$\alpha$ EqW-age relationship in SB99. 
 We estimate the age of the star forming region is 8 $\times$ 10$^{6}$ years, with a range of
  of $1 \times 10^{6}$ years permitted by the variation in the model and our uncertainty in $EW_{H\alpha}$.
 A similar estimate was obtained using the equivalent width of the H$\beta$ line, $EW_{H\alpha}=50\AA$.

\smallskip
\smallskip
\noindent{\sc Mass:}~
 The mass of a star forming region can be estimated by comparing the absolute $V$ or $B$ broadband
  magnitude to the magnitude-age trend in SB99, given for $M_V$ in the bottom left panel of Figure~\ref{fig:sb99}.
 The SB99 models assume a mass of 10$^{6}$ M$_{\odot}$, but the total broadband flux scales linearly
  with the mass of the region. 
 Given our de-reddened magnitude, $M_V$ = -15.76, we estimate a total stellar mass of $\sim 2 \times 10^{7} M_{\odot}$.
 For consistency, we also estimate the mass using the $M_B$ magnitude to obtain a similar estimate. 

\smallskip
\smallskip
\noindent{\sc Comparing other Parameters:}~
 Lastly, to confirm that our SB99 estimated parameters are all self-consistent, 
  we re-normalize the SB99 model to our assumed mass and compare to
  other physical parameters (Figure~\ref{fig:sb99}).
 First, we ensure that the number of Lyman continuum photons is consistent 
  with the mass of our region.
 As shown in the top right panel of Figure~\ref{fig:sb99}, our measured N$_{Lyc}$
  is consistent with the SB99 model at our inferred stellar population age.
 Second, we compare the $(B-V)$ color to the age in the bottom right panel of Figure~\ref{fig:sb99}. 
 The de-reddened color of our region, $(B-V)_{o}$ = 0.05 is consistent with the color of 
  the model at this age, given a conservative error in the color and the de-reddening 
  (shown as the color range in bottom right panel of Figure~\ref{fig:sb99}). 
 Given the overall consistency between our suite of observational data and SB99, 
  we are confident that, given our data, these parameters provide a reasonable description of 
  the blue overdensity.
 The adopted physical parameters for the star formation region are 
  summarized in Table 2.

\subsection{Discussion of Star Forming Properties}\label{sec:sfmode}
A component to understanding the origins of the blue overdensity in NGC\,5387 
 is to characterize the mode of star formation in the blue overdensity. 
We will do this by comparing our measurements for the blue overdensity 
 to those of a number of well studied star formation regions. 

\begin{figure}[tb]
\plotone{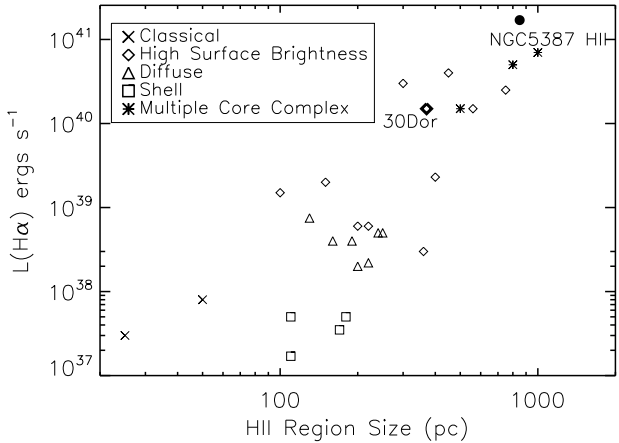}
\caption{\label{fig:hiireg} The observed size - H$\alpha$ luminosity - type relationship for the sample of 
 HII regions considered in \citet{kennicutt1984}. 
The regions are coded by the type of HII region: multiple core complexes as stars, 
 shell as squares, diffuse as triangles, high surface brightness as diamonds and classical as xes.
The HII region in NGC\,5387 is both physically larger and more luminous than the classical giant HII 
 regions 30 Doradus in the LMC and N604 in M33. 
Instead, it has properties similar to the ``multiple core complex'' HII regions like N5471, N5461 and N5455 in M101 (stars),
  implying the blue overdensity is composed of multiple smaller star forming regions.
}
\end{figure}

First, we evaluate if the size and luminosity of the region are more consistent with a single H\,II region or 
 a larger complex. 
In Figure~\ref{fig:hiireg}, the size and H$\alpha$ luminosity of the NGC\,5387 H\,II region are compared to the sample of
  \citet{kennicutt1984}; this demonstrates that the blue overdensity is both larger and more luminous than single 
 HII regions. 
Instead, the NGC\,5387 H\,II region has both a size and luminosity characteristic of ``multiple H\,II region complexes'' 
 observed in giant spirals like M\,101. 
As discussed by \citet{kennicutt1984}, these regions could contain several individual 
 30 Doradus-class H\,II regions in a single symmetric envelope of ionized and neutral gas.
While these star forming complexes are generally seen in large galaxies, 
 they are also observed in star-forming dwarf galaxies, 
 like NGC\,4449, M\,33 and the LMC \citep{kennicutt1984}. 
Thus, we move forward considering the blue overdensity is composed of multiple, unresolved regions
 that are forming stars. 

Following the example of \citet{Hunter1999}, diagnosis of the mode of star formation is akin to understanding the
 concentration stars within an H\,II region.
To demonstrate, we compare 30 Doradus in the LMC and Constellation III in the Milky Way, two regions
 that have similar mean properties.
The 30 Doradus region contains the super-star cluster R\,136, a $\sim$2 pc sub-component that is the origin of 
 most of the flux from the entire complex. 
Constellation III, in contrast, is described as a ``scaled up OB association'' --- a large, but overall low density
 region of young stars.
Constellation III and 30 Doradus have similar sizes and fluxes, but represent drastically different 
 star forming environments. 
The observational difference between these regions is the density of high mass stars, though the 
 comparison is complicated due to slight difference in age (30 Doradus $\lessapprox$ 8 Myr, 
 and Constellation III $\sim$ 10 Myr). 

To characterize the star forming region in NGC\,5387, we can make a number of comparisons. 
First, we can compare the blue overdensity to 30 Doradus, for which
 the total extent is $R\sim100$pc and the total H$\alpha$ luminosity is $L$(H$\alpha$)= 3.2$\times$10$^{39}$ \citep{lopez_2013}. 
The blue overdensity would have to contain 54 individual 30 Doradus-like regions to produce is measured
 H$\alpha$ luminosity (see Table 2) and by volume it could contain 600 such regions (assuming spherical geometry).
Second, we can compare to the ensemble of super star cluster candidates in NGC\,4449 identified by \citet[Table 5]{reines2008}.
The sample of 12 super star cluster candidates comprise a total mass of 2$\times$10$^{5}$ M$_{\odot}$,
 and produces N$_{LyC}$ = 3.7$\times$10$^{51}$ ionizing photons.
Scaled to our total mass, 2$\times$10$^{7}$ M$_{\odot}$, our ionizing flux is 
 comparable to a scaled up version of those regions in NGC\,4449.

\section{INTERPRETATION OF THE OBSERVATIONAL DATA}\label{sec:interp}
We have presented a set of observations that confirm the existence of a stellar stream in the halo of NGC\,5387 
 that, in projection, intersects the disk of NGC\,5387 near a prominent blue overdensity.
There are no distinguishable overdensities within the angular extent of the stream
 and we do not directly identify a progenitor within the stream itself.
Spectroscopy of the blue overdensity confirms it is associated with NGC\,5387.
Derived properties from nebular emission lines indicate it to be a young, massive, metal-poor H\,II complex. 
These observations are consistent with a number of physical scenarios. 

\subsection{Star Formation in the Disk?}
It is possible that the blue overdensity represents star formation occurring naturally in the disk,
 as is the case for a very massive, luminous and young star cluster in the spiral galaxy NGC\,6946 studied by \citet{Larsen2001}.
The existence of this cluster is an example of how ``violent interactions'' like mergers are not a required condition 
 to create ``extreme'' star forming environments in otherwise undisturbed galaxies \citep{Larsen2001}. 
The cluster in NGC\,6846, however, has a mass of 8.2$\times$10$^{5}$ M$_{\odot}$, much less massive than the blue overdensity in NGC\,5387. 
Furthermore, the SFR in the blue overdensity measured from H$\alpha$ 
 is nearly five times that measured in the center of NGC\,5387, making it a quite extreme event for the outer disk 
 of a MW analogue galaxy. 
Given that $\sim$100 of the clusters studied by \citet{Larsen2001} would have to exist in an 850 pc region, 
 this blue overdensity in NGC\,5387 seems to be a much more ``extreme'' case than that of NGC\,6946.
Thus, it seems unlikely that we are seeing star formation in the disk following an in-situ  mode. 

Having ruled out natural star formation in the NGC\,5387 disk, 
 it is possible that the blue overdensity is induced star formation. 
Generally, it is not clear what occurs when a dwarf satellite passes through the outer disk of its parent.
Current insight is limited to modelling that neither provides the proper spatial resolution,
 nor fully implements star formation prescriptions at the level needed to distinguish
 if the disk is ``transparent'' to the stream crossing or if there is sufficient 
 gravitational influence to induce star formation.
The observation of enhanced star formation in the outer disk of NGC\,5387 provides 
 one of the first observational test cases to explore the effects 
 of a minor merger on the outer disk of a Milky-Way analogue.
To evaluate this scenario, we can explore two key questions in our observational data: 
 (i) Are the kinematics of the blue overdensity consistent with the those of the NGC\,5387 disk?
 (ii) Is the metallicity of the gas consistent with that of the NGC\,5387 disk?

First, using $V_c$ from the NGC\,5387 HI profile \citep{h1profile}, we can estimate  
 the average rotational velocity of the disk at the radius of the overdensity.
Using $R_{sfr}$ = 8.7 to 9.1 kpc and the form of a standard rotation curve,
 the estimated disk motion in an edge-on galaxy should be near $V_{max}$, 
 which for NGC\,5387 is $V_{max}$ = 162.9 $\pm$ 3 km s$^{-1}$ \citep{h1profile}.
Given our uncertainty in the actual 3-dimensional position in the disk due to the inclination ($i = 77^{\circ}$),
 we could expect to see deviation from V$_{max}$ due to non line-of-sight motions.
With a velocity offset of $\delta$v = 104 km s$^{-1}$, 
 the blue overdensity has kinematics not inconsistent with being in the disk of NGC\,5387, 
 though with a relatively large dispersion.

For a basic estimation of the metallicity-radius relationship in the NGC\,5387 disk, 
 we employ the HII region metallicity-radius relationship observed in the Milky Way by \citet{balser2011}. 
We use their $360^{\circ}$ relation, 
\begin{equation}
  12+\log(O/H) = 8.866 - 0.0383 \times R_{gal}, 
\end{equation}
 and renormalize using the metallicity of the central region of NGC\,5387 from 
 SDSS spectroscopy, $12 +\log(O/H) = 9.05$ \citep{Tremonti2004}.
 we estimate that at the location of the HII region, $R_{sfr}$= 8.7 kpc, 
This implies a metallicity of $12+\log(O/H) =  8.73$ at the galactocentric radius of the blue overdensity. 
The NGC\,5387 HII region ($12+\log(O/H) = 8.03$) is more metal poor than that expected in the native gas 
 at this radius.  

However, taken together it is not impossible that star formation has been induced in the outer disk of NGC\,5387.
We must pose several questions relating to this hypothesis.
First, can such an event occur in the outer disk of a typical spiral galaxy?
Second, can we explain the deviations from typical kinematic and chemical abundance 
 trends in a spiral galaxy?

 \subsection{Tidal Dwarf Hypothesis?}

 Third, the blue overdensity could be formed from a gaseous component of the stellar stream intersecting the disk,
  or from the collision of a high velocity cloud (HVC) with the stream itself,
  though either of these scenarios seems highly unlikely.
 The formation of such a star forming region is analogous to the formation 
  of a tidal dwarf in a major merger, though on a much smaller and less extreme scale.  
 This ``tidal dwarf'' interpretation relies on there being an expelled gaseous component 
  to this minor merger or for there to be considerable gas reservoirs in the halo of NGC\,5387.
  
 The HI content of NGC\,5387 can be explored using the Arecibo 305m
  HI line profile of \citet{h1profile}.
 The beam size of Arecibo, 189$''$, contains the full disk of NGC\,5387, including
  the full stellar stream of NGC\,5387.
 The profile shows no significant distortions or other abnormal features to suggest
  that NGC\,5387 has any significant HI content outside of the expected disk motion.
 The profile, however, has an average channel width of 11 km s$^{-1}$ and an average rms of 1.67 Jy \citep{h1profile}, corresponding
  to a mass resolution of 1.3 $\times$ 10$^9$ $M_{\odot}$ at 80 Mpc in a single 11 km s$^{-1}$ channel.
 This sensitivity may not detect a gaseous stream or large cloud.
 Thus, while we cannot rule out that the NGC\,5387 stream has a gaseous component,
  there is no clear observational evidence from archival data for any abnormal HI features in NGC\,5387.
 
 On the other hand, the observed population of high velocity clouds (HVCs) in the Milky Way and Andromeda 
  do suggest that there can be relatively large clouds of gas moving within the halo,
  and such gas clouds are considered a normal component of MW-sized galaxy \citep[See ][ and references therein]{putman2012}.
 The Milky Way HVC population includes clouds with total gas masses greater than 10$^{6} M_{\odot}$, 
  typical physical sizes of $\sim$1-2 kpc, and estimated metallicities $\sim$0.1 Z$_{\odot}$ \citep{wakker2001}. 
 The total mass of even the largest of the HVCs is an order of magnitude smaller than the 
  stellar mass of the NGC\,5387 blue overdensity, 2 $\times$ 10$^7$ M$_{\odot}$.
 Additionally, the high FUV SFR indicates that the region has been forming stars for 
  an extended period of time, 100 Myr or more, 
  at which point an HVC would have already extinguished its gas reservoir \citep{LVLSF}.
 Thus, it seems unlikely that such a star forming region could be generated from an HVC collision with the stream.

\subsection{Stream Progenitor Hypothesis?}

The blue overdensity could be the progenitor of the stream. 
More specifically, it is possible that we have caught the satellite galaxy soon after its passage through the disk, 
 but it is the satellite itself, not the disk, that is currently undergoing star formation. 
Given that progenitors of the stream should be detected along the orbit of the stream,
 the blue overdensity is a strong candidate for the progenitor.
Typically, when progenitors can be connected to a stellar streams, those progenitors are already gas poor 
 and contain no appreciable on-going star formation or very young ($<$ 1 Gyr) stellar content,
 though we do note that there is a 2 Gyr population in the Sagittarius dSph \citep{siegel2007}.
Thus, for most observed stream progenitors there is little to no gradation in color 
 along the stream; the progenitor and the stream are largely indistinguishable in color.
In fact, in a search for MW-LMC analogue systems (within a 60 Mpc volume),
 many systems have been identified with long stellar streams or with tidal distortions 
 in the satellite, but none have exhibited strong color gradients 
 (Mart{\'{\i}}nez-Delgado et al., in prep.).
If the blue overdensity is the progenitor of the stream, then this is the first minor merger
 to be detected with a strong a color gradient. 

A clear morphological analogue to NGC\,5387 is the NGC\,5899 system, which also contains a red
 stellar stream in its outer disk.
The progenitor of the stream in NGC\,5899 is identified as a highly distorted overdensity along
 the stream near the inferred apo-centric radius (Mart{\'{\i}}nez-Delgado et al., in prep.). 
Because a satellite spends a large fraction of its orbit at apo-center, 
 most satellites are discovered at or near their apo-centric radii, 
 typically far from the disk of the parent galaxy.
The blue overdensity, however, is located in front of the NGC\,5387 disk in projection. 
If the blue overdensity is the progenitor, it is likely detected at or near its pericentric radius, 
 its closet approach to the NGC\,5387 disk, i.e., an unlikely position. 

Many dwarf spheroidal satellites in the Local Group show periodic episodes
 of enhanced star formation with a cadence similar to their orbital parameters \citep{hstphot}.
It is, however, exceedingly rare to find a star forming satellite in the halo of its parent \citep{geha2012},
 to find any significant gas reservoirs in satellite galaxies \citep{gp2009}, 
 or to find close satellites with colors similar to the LMC \citep{t11}.
Taken together, the relative scarcity of star forming dwarf satellites implies that the removal of gas 
 from satellite galaxies is highly efficient, 
 occurring very quickly relative to the survival timescale of the progenitor.
If the blue overdensity is the progenitor for the stream, 
 then it, as a dwarf satellite, presents exceptionally rare characteristics.

One of the properties of the blue overdensity is the discrepancy between the SFR estimated from 
 the H$\alpha$ and FUV luminosity. 
The H$\alpha$ luminosity is most sensitive to the highest mass
 stars whose lifetimes are very short, 
 and therefore this SFR indicator traces the most recent star formation.
In contrast, the FUV SFR traces star formation of 100 Myr,
 and the estimated FUV SFR implies that the blue overdensity has been forming stars 
 for at least the past 100 Myr, a timescale smaller than the inferred timescale for the formation of the stream.
The typical SFR for a dwarf galaxy of the same mass, traced by the total $B$ magnitude M$_{B}$,
 is 0.03 M$_{\odot}$ yr$^{-1}$, as estimated from the Local Volume Legacy Survey\citep{kennicutt2008}.
Thus, the both the $L(H\alpha)$ and $L(FUV)$ SFR measurements of the blue overdensity
 are enhanced compared to similar dwarf systems. 
The orbit for progenitor of the stream could have led to prior star formation enhancements 
 induced by the merging process. 
In addition to being consistent with timescales implied from the SFR indicators, 
 the stream progenitor hypothesis is also consistent with the low metallicity measured for the blue overdensity,
 and, given the mass of the stream debris, the small mass of the region. 

The stream progenitor origin of the blue overdensity is compelling 
 because this interpretation is consistent with both the physical parameters of the
 blue overdensity and from the merger properties inferred from the stellar stream. 
If the blue overdensity is the progenitor of the stream, then we are observing a 
 galaxy undergoing a period of significantly enhanced star formation during a minor merger.
Although the observational evidence for some enhancement of star formation in such events is inferred
 form the star formation histories of Local Group galaxies \citep{hstphot}, such a large enhancement has not 
 been previously observed for a minor merger nor has such an enhancement been predicted in numerical simulations.

Thus, we must pose several questions relating to this hypothesis.
First, can a dwarf galaxy maintain its gas reservoirs through a merger having a relatively small pericentric radii?
Second, can the dwarf form stars at the measured SFR and over the time scales implied by the FUV SFR? 

\subsection{Chance Alignment?}

Lastly, it must be considered that the stellar stream and the blue overdensity are a chance alignment 
 --- that the physical origins of the stream and the blue overdensity are unrelated. 
It is clear that the stellar stream is an accretion remnant, 
 and, given its redshift, the blue overdensity would have to be a second 
 satellite of dwarf Irregular type within NGC\,5387 system. 
This two satellite scenario, however, is a very low probability event.
First, the NGC\,5387 stream has a relatively high surface brightness, 24.5 mag arcsec$^{-2}$, 
 compared both to the known sample of stellar streams \citep{streamsurveypaper,mbd2011}
 and to those streams in the simulated halos of \citet{bj2005}.
Second, star forming satellites within close to their host are exceptionally rare.
In a sample of close host-satellite pairs in SDSS, \citet{t11} found
 only 10\% of MW analogues had a satellite within 50 kpc, and of those
 only 20\% of those are bluer than M33, 1.3\% bluer than the LMC.
NGC\,5387 would be accreting two of its largest satellites within $\sim$1 Gyr of each other
 and both of those systems are exceptional in their properties.
Thus, we find it statistically unlikely that the blue overdensity is 
 independent of the stellar stream given the statistics, 
 both observed and simulated, on such events individually.

\subsection{Discussion of Interpretations}

In this Section, we have considered four interpretations of our observational data on the blue overdensity of NGC\,5387: 
 (i) disk star formation (induced or in situ), (ii) a tidal dwarf analogue system, 
 (iii) a star burst in the stream progenitor, and (iv) a chance alignment of two unrelated accretion events.
Using our observational evidence and modeling NGC\,5387 based on radial trends in the Milky Way, 
 we have found that both the tidal dwarf and the chance alignment scenarios are highly unlikely.
The remaining two scenarios, merger-induced star formation either in the parent disk or the stream progenitor,
 while better fits to our observational data, are physical situations that have 
 neither been directly observed, nor modelled sufficiently to inform our interpretation. 
It remains unclear if the passage of a satellite galaxy through the outer region 
 of a parent disk could induce star formation of this degree ---
 even though there are examples of satellites in the Local Group that may have passed through 
 their parent's disk in the recent past \citep{barmodel,2massSgr}.
To further understand our observations, we require numerical modeling
 to determine if either of our two preferred scenarios are feasible.
These simulations and a further exploration of the NGC\,5387 system
 will be presented in a companion paper. 

\section{Summary}

Our deep VATT observations of NGC\,5387 show a stellar stream in its halo that interests
 the NGC\,5387 disk near a blue overdensity. 
Remarkably, both features were first discovered in SDSS, 
 showing that streams of a 1:50 mass ratio can be visually identified in SDSS, 
 even where other search techniques fail \citep{mbd2011}.
The measured properties of NGC\,5387, the stellar stream, 
 and the blue overdensity are given in Tables 1 and 2. 

The stellar stream has a median surface brightness of $R\sim$24.5 mag arcsec$^{-2}$ 
 and a total magnitude, $R$ = 17.5. 
The stream subtends 360$^{\circ}$ and wraps around the disk of NGC\,5387, with a width of 11$''$ (4.2 kpc).
Though there are SB variations, the stream appears smooth overall.
Lower surface brightness wraps of the stream are ruled out to our surface brightness
 sensitivity limit of 28.3 mag arcsec$^{-2}$.
The stream is red in color, $(B-V)\sim$0.70, 
 suggestive that it is comprised primarily of old stellar populations. 
The total stellar mass in the stream is 6$\times$10$^{8}$ M$_{\odot}$ and the stream
 appears to have formed in $\sim$400 Myr from a progenitor with a total mass 2$\times$10$^{10}$ M$_{\odot}$.

In contrast to the stellar stream, the blue overdensity is 
 blue in color, $(B-V) \sim$ 0.30, and FUV bright, implying it is actively forming stars.
While the blue overdensity can be associated with the NGC\,5387 system, 
 we cannot confidently associate it with the stellar stream.
If it is associated with the stream, however, this is the largest color
 gradient observed in accretion debris \citep{streamsurveypaper}
 and implies that it is possible for a satellite core to 
 maintain its gas at least $\sim$400 Myr into the merging process.
Numerical simulations to explore the relationship between these two features
 will be presented in a companion paper. 

While the relationship between the overdensity and the stellar stream remains indeterminate,
 our classification of the blue overdensity as a star forming region is clear.
The blue overdensity is $\sim$850 pc in diameter, metal poor ($12+\log(O/H)$=8.03), 
 and has a star formation rate of 1.19 M$_{\odot}$ yr$^{-1}$ from $L$(H$\alpha$)
 or 2.72 M$_{\odot}$ yr$^{-1}$ from $L$($FUV$). 
The stellar mass for the region is 2$\times$10$^{7}$ M$_{\odot}$
 and the age of the stellar population is $\sim$ 8 Myr,
 though the $L$($FUV$) implies the region has been forming stars for at least 100 Myr \citep{LVLSF}.

A comparison of the blue overdensity to H\,II regions in the nearby Universe (Figure 6) implies that 
 it is most likely composed of several individual H\,II regions
 that we are unable to distinguish due to the seeing (1$''$) and 
 the VATT spatial resolution (0.38$''$ per pixel) of current imaging data.
Furthermore, the region is producing ionizing photons comparable to 
 that produced by the entire super star cluster population in NGC\,4449 
 (scaling to total mass), i.e., a dwarf galaxy forming stars at twice the rate of the LMC \citep{reines2008}.
Thus, we infer that the star formation mode in this region is akin
 to those modes observed in nearby star-burst dwarf irregular galaxies.

Combining all of our observational data, there are four possible interpretations:
 (i) the blue overdensity represents star formation induced in the disk by the minor merger,
 (ii) the blue overdensity is a dwarf formed during the interaction,
 (iii) the blue overdensity is the galaxy remnant of the progenitor of the stream in a burst phase, 
 and (iv) the blue overdensity is an unrelated object. 
After careful consideration, we reject both the tidal dwarf (Section 4.2)
 and the chance alignment scenarios (Section 4.4).
Thus, the blue overdensity most likely represents star formation induced by
 a minor merger -- either in the disk of the parent or in the progenitor of the stream. 
Neither phenomenon has been directly observed previously for an accretion event of this size.
Given the similarities of the blue overdensity to NGC\,4449 and 
 the enhanced SFR in FUV (sensitive to SF of age $\sim$100 Myr),
 the blue overdensity is most likely the progenitor of the stream caught in a burst phase.
Confirmation of this interpretation via detailed simulations
 and a full exploration of the implications of the NGC\,5387 system will be presented in a companion paper.

\acknowledgements
We acknowledge assistance from R. Pogge in planning, collecting and reducing MODS data.
We acknowledge helpful conversations with Robert O'Connell, Trihn Thuan, Sabrina Stierwalt, Amy Reines and David Whelan.
RLB recognizes the Jefferson Scholars Foundation and served as the Mark C. Pirrung Family Fellow in Astronomy,
 as well as recognizing the Office of the Vice-President for Research at the University of Virginia.
RLB and SRM recognize support from NSF grants AST-1009882 and AST-0607726.
ED gratefully acknowledges the support of the Alfred P. Sloan Foundation.
 This paper used data obtained with the MODS spectrograph built with
  funding from NSF grant AST-9987045 and the NSF Telescope System
  Instrumentation Program (TSIP), with additional funds from the Ohio
  Board of Regents and the Ohio State University Office of Research.
 Funding for the SDSS and SDSS-II has been provided by the Alfred P. Sloan Foundation, the Participating          
Institutions, the National Science Foundation, the U.S. Department of Energy, 
the National Aeronautics and Space Administration, the Japanese Monbukagakusho, the Max Planck Society, 
and the Higher Education Funding Council for England. The SDSS Web Site is http://www.sdss.org/.
This research has made use of the NASA/IPAC Extragalactic Database (NED) which is operated by the Jet Propulsion Laboratory, California Institute of Technology, under contract with the National Aeronautics and Space Administration.
Funding for SDSS-III has been provided by the Alfred P. Sloan Foundation, the Participating Institutions, the National Science Foundation, and the U.S. Department of Energy Office of Science. The SDSS-III web site is http://www.sdss3.org/.
SDSS-III is managed by the Astrophysical Research Consortium for the Participating Institutions of the SDSS-III Collaboration including the University of Arizona, the Brazilian Participation Group, Brookhaven National Laboratory, Carnegie Mellon University, University of Florida, the French Participation Group, the German Participation Group, Harvard University, the Instituto de Astrofisica de Canarias, the Michigan State/Notre Dame/JINA Participation Group, Johns Hopkins University, Lawrence Berkeley National Laboratory, Max Planck Institute for Astrophysics, Max Planck Institute for Extraterrestrial Physics, New Mexico State University, New York University, Ohio State University, Pennsylvania State University, University of Portsmouth, Princeton University, the Spanish Participation Group, University of Tokyo, University of Utah, Vanderbilt University, University of Virginia, University of Washington, and Yale University.

\include{table}

\bibliographystyle{apj}
\bibliography{bib}

\end{document}

%% file: table.tex
\begin{deluxetable}{ccccc}
\tablecaption{Photometric Properties of the NGC\,5387 System}
\label{tbl:phot}
\tablehead{ 
  \colhead{Photometry Property} & \colhead{NGC\,5387} & \colhead{Stellar} & \multicolumn{2}{c}{Blue Overdensity} \\
  \colhead{ }                   & \colhead{Total}     & \colhead{Stream}  & \colhead{E($B-V$)=0.03} & \colhead{E($B-V$)=0.35} 
 }
\startdata
 $M_{FUV}$  & -17.06  (1)& \ldots      & -16.39  & -20.83  \\
 $M_{NUV}$  & -17.67  (1)& \ldots      & -16.53  & -19.36  \\
 $M_{B}$    &  14.46     &  $\sim$19   & -14.42  & -15.72  \\
 $M_{V}$    &  13.67     &  $\sim$18.1 & -14.74  & -15.76  \\
 $M_{R}$    &  13.10     &  17.5       & -14.89  & -15.67  \\
 ($B-V$)$_{o}$ & 0.79 & 0.6-0.9        & 0.32  &   0.04  \\
 ($B-R$)$_{o}$ & 1.36 & 1.2-1.7        & 0.47  &  -0.05  \\
 $M_{*}$   & 3.13$\times$10$^{10}$ M$_{\odot}$ &  \ldots & \ldots & \ldots \\
 $M_{tot}$ & 1.1$\times$10$^{11}$ M$_{\odot}$  & 2$\times$10$^{10}$ M$_{\odot}$ & \ldots & \ldots 
\enddata
\tablenotetext{}{{\it References:} (1) \citet{lemonias2011} }
\end{deluxetable}

\begin{deluxetable}{cccc}
\tablecaption{Spectroscopic Properties of the NGC\,5387 System}
\label{tbl:spec}
\tablehead{
 \colhead{Property} & \colhead{NGC\,5387} & \multicolumn{2}{c}{Blue Overdensity} \\
 \colhead{ }        & \colhead{Total}     & \colhead{E($B-V$)=0.03} & \colhead{E($B-V$)=0.35}
}
\startdata
 $v_{los}$         & 5226 km $\pm$ 3 kms$^{-1}$         & 5331 $\pm$ 2 kms$^{-1}$ & 5331 $\pm$ 2 kms$^{-1}$  \\
 $M_{*}$           & 3$\times$10$^{10}$ M$_{\odot}$ (1) &        & 2$\times$10$^{7}$ M$_{\odot}$  \\
 $12+\log(O/H)$    & 9.05 (2)                           & 8.04   & 8.03   \\
 Age               & \ldots                             & 8 Myr  & 8 Myr \\
 $L$(H$\alpha$)      & \ldots                             & 9.97$\times$10$^{40}$ ergs s$^{-1}$   & 2.24$\times$10$^{41}$ ergs s$^{-1}$ \\
 SFR $L$(H$\alpha$)  & 0.22 M$_{\odot}$ yr$^{-1}$ (3)     & 0.53 M$_{\odot}$ yr$^{-1}$            & 1.19 M$_{\odot}$ yr$^{-1}$     \\
 $L$(FUV)            & \ldots                             & 3.0$\times$10$^{26}$ ergs s$^{-1}$   & 1.95$\times$10$^{28}$ ergs s$^{-1}$ \\
 SFR $L$(FUV)        & \ldots                             & 0.04 M$_{\odot}$ yr$^{-1}$            & 2.72 M$_{\odot}$ yr$^{-1}$     \\
 $N_{LyC}$         & \ldots                             & 7.83 $\times$10$^{52}$ photons        & 1.35$\times$10$^{53}$ photons   
\enddata
\tablenotetext{}{{\it References:} (1) \citet{Kauffmann2003}, (2) \citet{Tremonti2004}, (3) \citet{Brinchmann2004} for the central region only}
\end{deluxetable}